\documentclass[twocolumn,twocolappendix,floatfix]{aastex701}

\usepackage{times}
\usepackage{amsmath}
\usepackage[varg]{txfonts}
\usepackage{bm}
\usepackage{threeparttable}
\usepackage{changepage}
\usepackage{booktabs}
\usepackage{multirow}   

\begin{document}

\title{Reconstructing the mass-redshift distribution of supermassive black holes \\using multiband gravitational-wave observations}

\author{Alorika Kar}
\affiliation{International Centre for Theoretical Sciences, Tata Institute of Fundamental Research, Bangalore, 560089}
\email[hide]{alorika.kar@icts.res.in}  

\author{Souvik Jana}
\affiliation{Department of Physics, The Chinese University of Hong Kong, Hong Kong, Shatin, China}
\affiliation{International Centre for Theoretical Sciences, Tata Institute of Fundamental Research, Bangalore, 560089}
\email[hide]{souvik.jana@icts.res.in}

\author{Enrico Barausse}
\affiliation{SISSA, Via Bonomea 265, 34136 Trieste, Italy}
\affiliation{INFN Sezione di Trieste, Via Valerio 2, 34127 Trieste, Italy}
\affiliation{IFPU - Institute for Fundamental Physics of the Universe, Via Beirut 2, 34014 Trieste, Italy}
\email[hide]{barausse@sissa.it}

\author{Parameswaran Ajith}
\affiliation{International Centre for Theoretical Sciences, Tata Institute of Fundamental Research, Bangalore, 560089}
\email[hide]{ajith@icts.res.in}  

\begin{abstract}

The formation and evolution of supermassive black holes (SMBHs) remain major open questions in astrophysics. Astrophysical models of SMBHs invoke black-hole (BH) seeds formed at high redshift, which subsequently grow through gas accretion and mergers. Although observations of high-redshift quasars and active galactic nuclei constrain these scenarios, the mechanisms governing SMBH formation and growth remain uncertain. Future gravitational-wave (GW) observations across the millihertz, decihertz, and audio-frequency bands will provide a complementary and direct probe of SMBH assembly by tracing their merger history. We investigate the extent to which multiband GW observations can reconstruct the mass-redshift distribution of these mergers.  We simulate populations of BH mergers predicted by different astrophysical models and identify those observable by the millihertz detector LISA, the decihertz detector LGWA, and next-generation ground-based detectors operating in the audio-frequency band. We reconstruct the mass-redshift distribution of these mergers using the nonparametric inference code \textsc{Figaro}. We find that combining observations across frequency bands substantially improves the reconstruction for the light-seed models considered here, whose merger populations span broad ranges of mass and redshift. In contrast, for the heavy-seed models considered here, LISA alone recovers most of the accessible mass–redshift distribution, with little additional improvement from the multiband network.

\end{abstract}

\section{Introduction} 

Gravitational-wave (GW) observations by LIGO~\citep{LIGO} and Virgo~\citep{Virgo} have opened a new window on the astrophysics of black holes (BHs). Historically, BHs have been observed mainly in two mass regimes: stellar-mass ($\sim 10-100 M_\odot$) BHs, formed as the remnants of massive stars, and supermassive ($\sim 10^6$--$10^{9}\,M_{\odot}$) BHs (SMBHs), found at the centers of most galaxies. Stellar-mass BHs were discovered as x-ray binaries in our own galaxy, while SMBHs were discovered through radio and optical observations of active galactic nuclei (AGNs) and quasars, out to redshifts $\sim 11$~\citep{jwst1, jwst2, Fan2006_highz_quasar, Banados2018_highz_quasar3, Wu2015_highz_quasar2, farthestzquasar2026}. Recently, pulsar timing arrays have obtained evidence of a stochastic GW background, thought to be produced by inspiralling binaries of SMBHs~\citep{PTA1, PTA2}. A new class of BHs, with intermediate masses ($\sim 10^2$--$10^5\,M_{\odot}$), is also postulated to exist. Until recently, observational evidence for these intermediate-mass BHs was scant. However, LIGO-Virgo observations suggest that some of the binary BH (BBH) mergers would have produced BHs with masses $\sim 100-200 M_\odot$ as remnants~\citep{highmassLVK1, highmassLVK2}. At the same time, massive BHs with masses $\sim 10^4-10^5 M_\odot$, which are lighter than the conventional SMBHs, are also being discovered in various galaxies~\citep{Baldassare_light_SMBH_2015, Greene_light_SMBHs_2004, jgreene_imbh_review, Reines_light_SMBHs, reinesdwarfgalaxy1}. The historical gap between stellar-mass and supermassive BHs appears to be shrinking.  

Despite compelling evidence for the existence of SMBHs, their formation and evolution remain difficult to explain within a single framework. The prevailing picture invokes BH seeds formed at high redshift, which subsequently grow through gas accretion and mergers. Light seeds (LSs), with characteristic masses of $\sim 100\,M_\odot$, may form from the remnants of metal-free Population III stars \citep{popIII2, bh_evolution3, johnson_popIII_ref, inayoshi_popIII_ref, tanaka_popIII_ref}. However, observations indicate that SMBHs with masses of $\sim 10^8\,M_\odot$ were already present at $z\sim7$--$9$ \citep{jwst1, jwst2, maiolino_jwst_observation, ubler_jwst_observation, li1_jwst_scaling, harikane_lrd, labbe_lrd, matthee_lrd, kocevski_lrd}. Growing such objects from LSs within the available cosmic time is challenging, although recent cosmological simulations suggest that sufficiently efficient gas accretion, together with realistic feedback, may allow LSs to account for at least part of the observed high-redshift SMBH population \citep{Mehta_popIII_jwst_reference, Madau_2014_rapid_bh_growth}.

{An alternative possibility is the formation of heavy seeds (HSs), with characteristic masses of $\sim10^4$--$10^6\,M_\odot$, through the rapid collapse of large gas concentrations in primordial or metal-poor protogalaxies \citep{Bromm:2002hb, Begelman:2006db, HS2, haiman_smstar, Begelman_armitage_smstar}. Depending on the thermodynamic evolution of the inflow, this process may pass through a supermassive-star or quasi-star phase. Massive seeds may also arise through runaway stellar collisions and gas-assisted growth in dense nuclear clusters. Although HSs alleviate the need for prolonged rapid growth when explaining high-redshift quasars, their formation generally requires restrictive conditions, including low metallicity, rapid gas inflow, suppressed fragmentation, and efficient angular-momentum transport \citep{Begelman:2006db, Lodato:2006hw, Bromm:2002hb, chon_2021_smstar_problem, Latif_review, Dijkstra_DCBH_paper, bhowmick_dcbh, habouzit_dcbh}. More exotic possibilities include primordial BHs formed from the collapse of large primordial density perturbations. Depending on their formation epoch and mass spectrum, these objects could provide either light or heavy seeds; however, their production is not a generic prediction of standard inflationary cosmology and requires particular early-Universe conditions \citep{carr_kuhnel_2020_review, carr_silk_review_2018}.}

The processes governing the growth of SMBHs from their seeds also remain poorly understood. To account for the most massive quasars observed at high redshift, LSs generally require prolonged accretion at close to (or above) the Eddington rate, with a high duty cycle \citep{Madau_2014_rapid_bh_growth}. Sustaining such flows (especially at super-Eddington rates) is challenging because the radiation produced by the accreting gas can oppose further accretion and drive powerful outflows. Feedback from AGNs and supernovae (SNe) may further regulate the supply of gas to the BH and interrupt its growth. BH mergers provide an additional growth channel, but their efficiency depends on the time required for the BHs to sink toward the galactic centre, form a bound binary, and harden to separations at which GW emission drives the final inspiral. In gas-poor and approximately spherical stellar systems, this evolution may stall at parsec scales, giving rise to the \emph{final-parsec problem} \citep{begelman_nature_1980, colpi_2014_tdelay_review, Milosavljevi_2001_final_parsec, Milosavljevi__2003_final_parsec, Dvorkin:2017vvm}. The observed SMBH population may contain contributions from multiple seed channels, shaped by environment-dependent accretion, feedback, and merger delays. Disentangling these contributions remains an open problem.

Recent observations with the James Webb Space Telescope (JWST) are refining our understanding of SMBH formation and early growth. JWST has identified SMBHs with masses  $\sim 10^{7}$--$10^{8}\,M_\odot$ at redshifts $z \gtrsim 8$, corresponding to epochs less than $\sim 700$ Myr after the Big Bang~\citep{jwst1, jwst2, harikane_lrd, ubler_jwst_observation, maiolino_jwst_observation, jades_lowmassmbh_highz}. Several of these systems appear to host BHs that are overmassive relative to their stellar hosts when compared with the local SMBH-galaxy scaling relations. This suggests that BH growth may have preceded the assembly of their host galaxies~\citep{Banados2018_highz_quasar3, harikane_lrd, maiolino_jwst_observation, ubler_jwst_observation}. Initial observations seemed to favor the HS models, as earlier studies had suggested that LSs are unable to grow into such overmassive SMBHs in such a short span of cosmic time. However, recent analyses of stacked JWST/JADES spectra have revealed evidence for a substantial population of lower-mass SMBHs ($\sim 10^{6}\,M_\odot$) at $3<z<7$, pointing to a more diverse early SMBH population than previously inferred \citep{jades_lowmassmbh_highz}. Fitting a parameterized BH population model jointly to PTA and JWST constraints points to heavier seeds, prompt BH coalescences after halo mergers, and phases of super-Eddington growth lasting tens of Myr \citep{Toubiana:2024bil}. Upcoming optical, X-ray, and radio facilities, including Euclid \citep{euclid1, euclid2}, the Vera C. Rubin Observatory \citep{lsst1, lsst2}, the Nancy Grace Roman Space Telescope \citep{nancy1, nancy2}, the Square Kilometre Array \citep{ska1, ska2}, and Athena \citep{athena1, athena2}, will further clarify these questions. 

Complementing these electromagnetic observations, ground- and space-based GW detectors will probe BH growth through BH mergers. {The physics of GW generation and propagation is well understood, and detector selection effects are much better modeled than those in electromagnetic observations}. The proposed next-generation (XG) ground-based observatories~\citep{CE,ET}, operating in the audio-frequency region ($\sim 5 - 10^3$ Hz) of the GW spectrum, will probe BBHs of mass $\sim 10-10^3 M_\odot$ out to redshifts $\sim 5-30$.  The upcoming space-based observatory LISA~\citep{amaroseoane2017laserinterferometerspaceantenna, lisa_universe}, operating in the millihertz frequencies ($\sim 10^{-4} - 10^{-1}$ Hz), will observe BBHs of mass $\sim 10^4-10^7 M_\odot$ out to redshifts $\sim {10-20}$. Several concepts have been proposed for space-based or lunar decihertz GW detectors: DECIGO~\citep{decigo}, LGWA~\citep{Harms_2021, lgwa}, LILA~\citep{lila}, IndIGO-D~\citep{Sharma:2026tff}, etc., operating in the $\sim 10^{-3} - 1$ Hz band that will detect BBHs of mass $\sim 10^3-10^5 M_\odot$. Depending on their sensitivity, these detectors could have an observable horizon up to redshifts of $\sim 10-30$. In summary, these observatories will observe BBH mergers of varying masses at different redshifts, essentially mapping the merger history of SMBHs over cosmic time, thus constraining models of SMBH growth. 

The potential of LISA observations to constrain the formation and evolution of SMBHs has been explored in several previous studies. Early forecasts suggested that the number of detected mergers and their distributions in mass, redshift, and mass ratio could distinguish between light- and heavy-seed scenarios and between different accretion prescriptions \citep{Sesana_2007}. \citet{plowman_2010_bh_mass_spectrum_1,plowman_2011_bh_mass_spectrum_2} investigated how LISA selection effects and parameter-estimation uncertainties affect the recovery of the BH mass spectrum and the comparison of population models. Bayesian frameworks for connecting catalogs of LISA detections to SMBH formation models were subsequently developed by \citet{gair_2011_lisa_population_constraint,Sesana_2011}, while \citet{mixture_fraction_enrico} used hierarchical Bayesian inference to constrain mixtures of competing seed and evolution scenarios. More recently, \citet{langen_hierarchical_bayesian} investigated how well LISA could constrain phenomenological parameters describing the BH--halo relation, occupation fraction, and delay between halo and BH mergers. Complementing these parametric studies, \citet{Sadiq_2025} presented a nonparametric reconstruction of the LISA BH-binary population using iterative kernel density estimation, explicitly accounting for measurement uncertainties and selection effects in an LS scenario. Recently, \citet{2026arXiv260500092I} compared merger populations predicted by a broad collection of semi-analytic models, quantifying the substantial theoretical variation in LISA event rates and mass-redshift distributions arising from different treatments of seeding, galaxy evolution, numerical resolution, and BH-merger delays.

A smaller number of studies have considered the complementary information provided by multiple GW frequency bands. \citet{vecchio_multiband_2009} estimated the detectability and parameter-estimation accuracy of LS BH mergers with LISA, DECIGO, and the Einstein Telescope. Using a semi-analytic model that follows light and heavy seeds in the progenitors of high-redshift quasars, \citet{valiante_multifrequency_bh_growth} showed that LS binaries may first become observable with ground-based detectors at high redshift and later enter the LISA band after substantial growth. \citet{Bhagwat:2021kwv} used semi-analytic LS and HS populations similar to those adopted here to assess ringdown spectroscopy with LISA and the Einstein Telescope, finding that the ringdowns of LS remnants partly fall in the ET band. More recently, \citet{davari_volonteri_2026_multiband_rates} used the CAT semi-analytic model to predict the merger and detection rates of binaries produced by light-, intermediate-, and heavy-seed scenarios for LISA, LGWA, and the Einstein Telescope.

{Building on these studies, we investigate how observations across the millihertz, decihertz, and audio-frequency bands can be combined to reconstruct the intrinsic mass-redshift distribution of merging BH binaries nonparametrically, including measurement uncertainties and detector selection effects, for several light- and heavy-seed models. We quantify the additional information provided by combining LISA with future decihertz and ground-based detectors.}  We simulate populations of merging BBHs over cosmic time as predicted by different models reported in~\cite{Barausse_2012, Kallol_PTA_implication, Barausse_2020} and compute their detectability in various detectors using a simple SNR criterion. For detectable BBHs, we create synthetic posterior distributions of the source parameters using a Fisher-matrix approximation. These synthetic posteriors of individual mergers are combined to reconstruct the astrophysical distribution of merging BBHs, after appropriately considering the selection function of various detectors. We perform this using the nonparametric Bayesian population-inference code \textsc{Figaro}~\citep{figaropackage, figaro2024}. We find that multiband observations significantly improve the reconstruction of the population properties of LS models. For the case of HS models, LISA is sufficient for accurate reconstruction.

The remainder of this paper is organized as follows. Section~\ref{sec:smbh_models} describes the astrophysical models considered in this study, while Sec.~\ref{sec:detectors} summarizes the GW detectors and their assumed sensitivities. Section~\ref{sec:reconstruct} presents our simulation and population-reconstruction framework, including the treatment of parameter-estimation uncertainties and selection effects. We present the reconstruction results in Sec.~\ref{sec:results} and summarize our conclusions, limitations, and directions for future work in Sec.~\ref{sec:summary}.

\section{Astrophysical models of SMBHs}
\label{sec:smbh_models}

\begin{figure*}[tbh]
\centering	
\includegraphics[width=2\columnwidth]{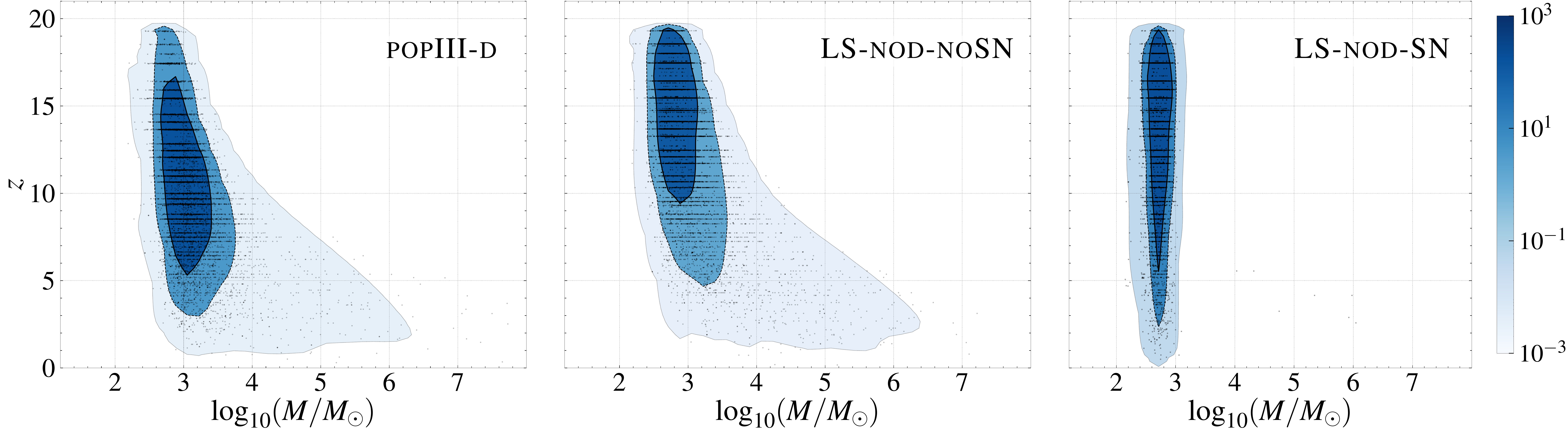}
\includegraphics[width=2\columnwidth]{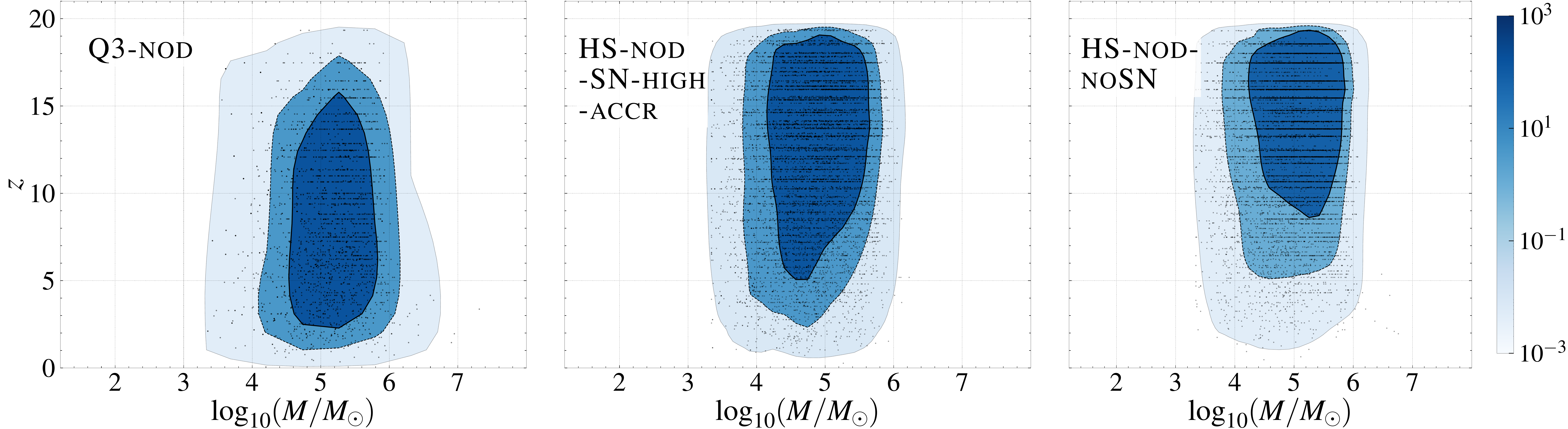}			
\caption{Intrinsic merger-rate density predicted by different astrophysical models used in our study. The top panel corresponds to the LS models, while the bottom panel corresponds to the HS models. The color map shows the observer-frame merger-rate density ${d^3N}/[{dt_{\rm o}\,d\log_{10}M\,dz}]$  in the corresponding mass-redshift bin.  The black points represent individual merger events used for our simulation, assuming an observation time of 10 years.}
\label{fig:merger_rate}
\end{figure*}

Most massive galaxies harbor a central SMBH, and strong correlations have been established between SMBH mass and host-galaxy properties, particularly the bulge stellar mass and stellar velocity dispersion \citep{Kormendy_2013, Ferrarese_2000, Barausse:2017uyr, sesana_m_sigma_ref, reines_m_sigma_ref}. These scaling relations suggest a close connection between the growth of SMBHs and the evolution of their host galaxies. Models of SMBH formation and evolution therefore generally include growth through gas accretion and repeated BH mergers, together with the processes governing the evolution of the host galaxy \citep{bh_evolution1, bh_evolution2, bh_evolution3, Barausse:2020kjy}. Most models assume that BH seeds with masses of  $\sim 10^2$--$10^5\,M_{\odot}$ form at high redshift. The formation and early growth of these seeds remain poorly constrained observationally.

In this work, we adopt the semi-analytic model introduced in \citet{Barausse_2012}, which jointly follows the formation and evolution of galaxies, their baryonic components, and their central BHs. The model was subsequently extended and refined in \citet{Sesana_2014, Antonini_2015, Bonetti_2018, Bonetti:2018tpf, Barausse_2020, Kallol_PTA_implication}. It is built on dark-matter halo merger trees generated using the extended Press-Schechter formalism \citep{press_schechter}. The baryonic components of galaxies evolve along these merger trees according to semi-analytic prescriptions. The model includes the evolution of the interstellar medium, star formation, SN and AGN feedback \citep{Barausse_2020}, and the formation and evolution of nuclear star clusters \citep{Antonini_2015, Antonini:2015cqa}. Below, we summarize the ingredients most relevant to the present analysis.

\paragraph{Seed BHs:}

The initial mass function and occupation fraction of BH seeds are important determinants of the subsequent BH merger population. We consider two representative scenarios: LSs and HSs. In the LS scenario, BH seeds form as remnants of Population III stars in metal-poor environments at high redshift \citep{bh_evolution3, popIII2, johnson_popIII_ref, inayoshi_popIII_ref, tanaka_popIII_ref}. The Population III progenitor masses are drawn from a log-normal distribution centered at $300\,M_{\odot}$ with a standard deviation of $0.2$ dex. Stars in the pair-instability range ($140\,M_{\odot}$ -- $260\,M_{\odot}$) are assumed to leave no BH remnant, while the remaining progenitors produce BH seeds with masses equal to approximately two-thirds of their stellar masses, accounting for mass loss during collapse.
In the HS scenario \citep{Begelman:2006db, HS2, Lodato:2006hw}, seeds form through the rapid collapse of gas in unstable protogalactic discs. Their masses depend on the properties of the host halo and are typically $\sim 10^5\,M_{\odot}$. 

\paragraph{Delay between galaxy and BH mergers:}

The predicted BH-binary merger rate depends strongly on the uncertain time delay between galaxy  and BH mergers. In the models used here \citep{Kallol_PTA_implication}, when two dark-matter haloes merge, the less massive halo becomes a satellite subhalo and sinks toward the center of the host through dynamical friction \citep{dyn_fric_chandrasekhar}. Tidal stripping and evaporation progressively reduce the mass of the satellite and can lengthen its sinking time.

Following the galaxy merger, dynamical friction due to stars and gas brings the two BHs from kiloparsec to parsec scales.  If the satellite galaxy is tidally disrupted before reaching the center, its BH may be left without a substantial stellar envelope. Such ``naked'' BHs can wander or stall at separations of tens to hundreds of parsecs \citep{binney_tremaine, Barausse_2020, tremmel_final_parsec}.

Once the two BHs become gravitationally bound, their subsequent evolution is driven by interactions with the surrounding stellar and gaseous environment \citep{colpi_2014_tdelay_review}. In gas-poor systems, three-body interactions with stars extract energy and angular momentum from the binary, ejecting stars and shrinking the orbit \citep{sesana_khan_three_body_int, quinlan_three_body_int}. If the supply of interacting stars is efficiently replenished, the binary can reach separations of approximately $10^{-3}$--$10^{-2}\,\mathrm{pc}$ within a few Gyr. At sufficiently small separations, GW emission dominates the orbital evolution and drives the binary to merger. The efficiency of these processes, and hence the total delay between the halo merger and the BH merger, depends sensitively on the structure and gas content of the  galaxy.

\paragraph{Accretion model}

Gas accretion is generally believed to be the dominant channel of BH growth. Because the inflowing gas carries angular momentum, it typically forms an accretion disc, through which angular-momentum transport allows matter to reach the BH. The Eddington luminosity is obtained by balancing the outward force exerted by radiation on ionized gas against the gravitational attraction of the BH. For a radiative efficiency $\epsilon$, the corresponding Eddington accretion rate is conventionally defined as $\dot{M}_{\rm Edd}=L_{\rm Edd}/(\epsilon c^2)$, while the BH grows at approximately $\dot{M}_{\rm BH}=(1-\epsilon)\dot{M}$. Accretion can exceed $\dot{M}_{\rm Edd}$ when photon trapping, advection, anisotropic radiation, and outflows reduce the effective radiative efficiency, enabling rapid BH growth \citep{paczynski_sup_edd, volonteri_sup_edd, wyithe_loeb_sup_edd}. Such super-Eddington episodes may be triggered by gas inflows following major halo mergers, defined in the model adopted here by $q_h=m_2^{\rm halo}/m_1^{\rm halo}\geq0.13$ \citep{ricarte_natarajan_major_merger}. Because LS models with strictly Eddington-limited growth have difficulty producing the SMBHs powering the most luminous quasars at high redshifts~\citep{Madau_2014_rapid_bh_growth}, the LS models considered here allow mildly super-Eddington accretion, up to twice the Eddington rate.

\begin{figure*}[t]
	\centering
	\includegraphics[height=2in]{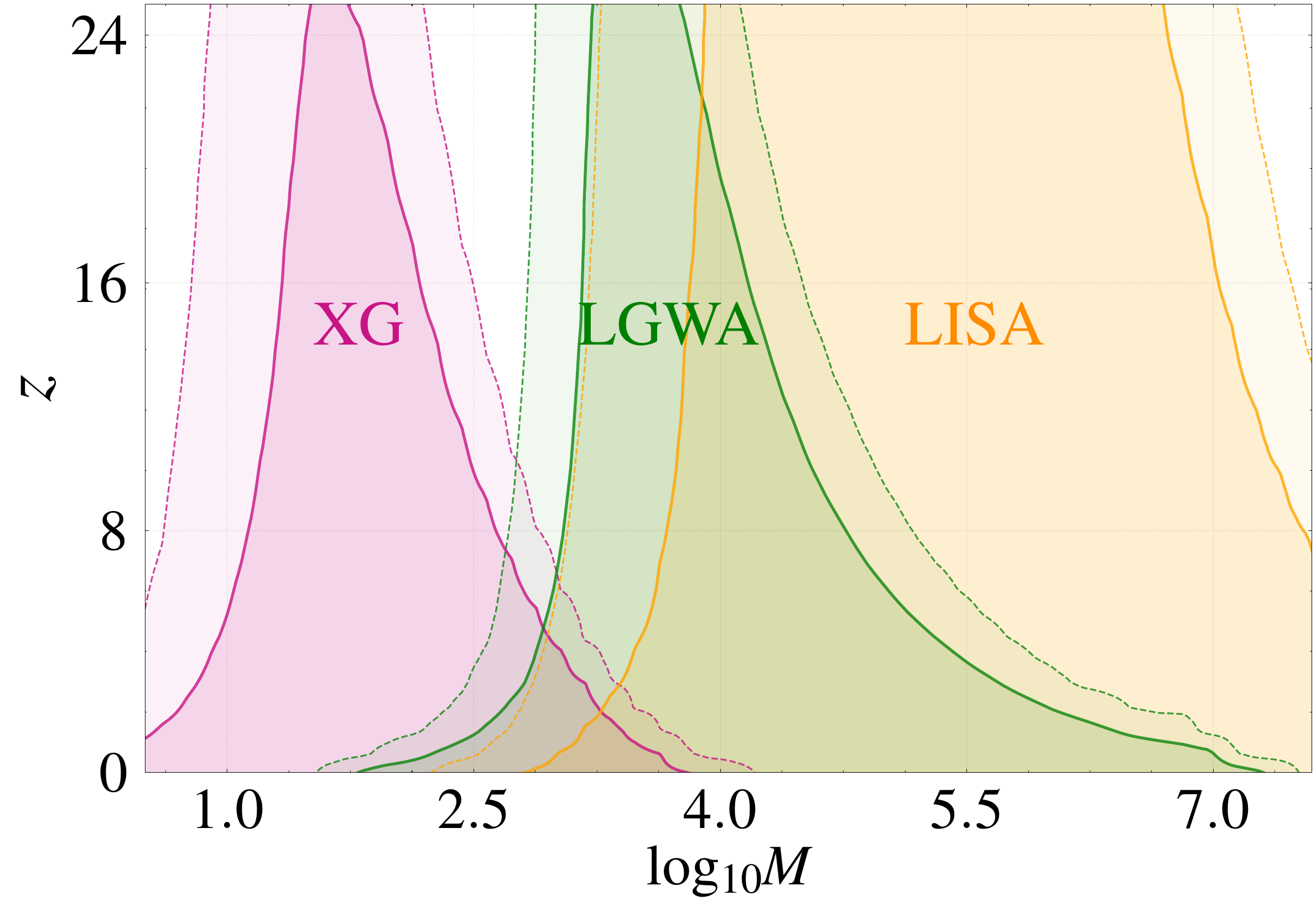}
	\hspace{0.02\columnwidth}
	\includegraphics[height=2in]{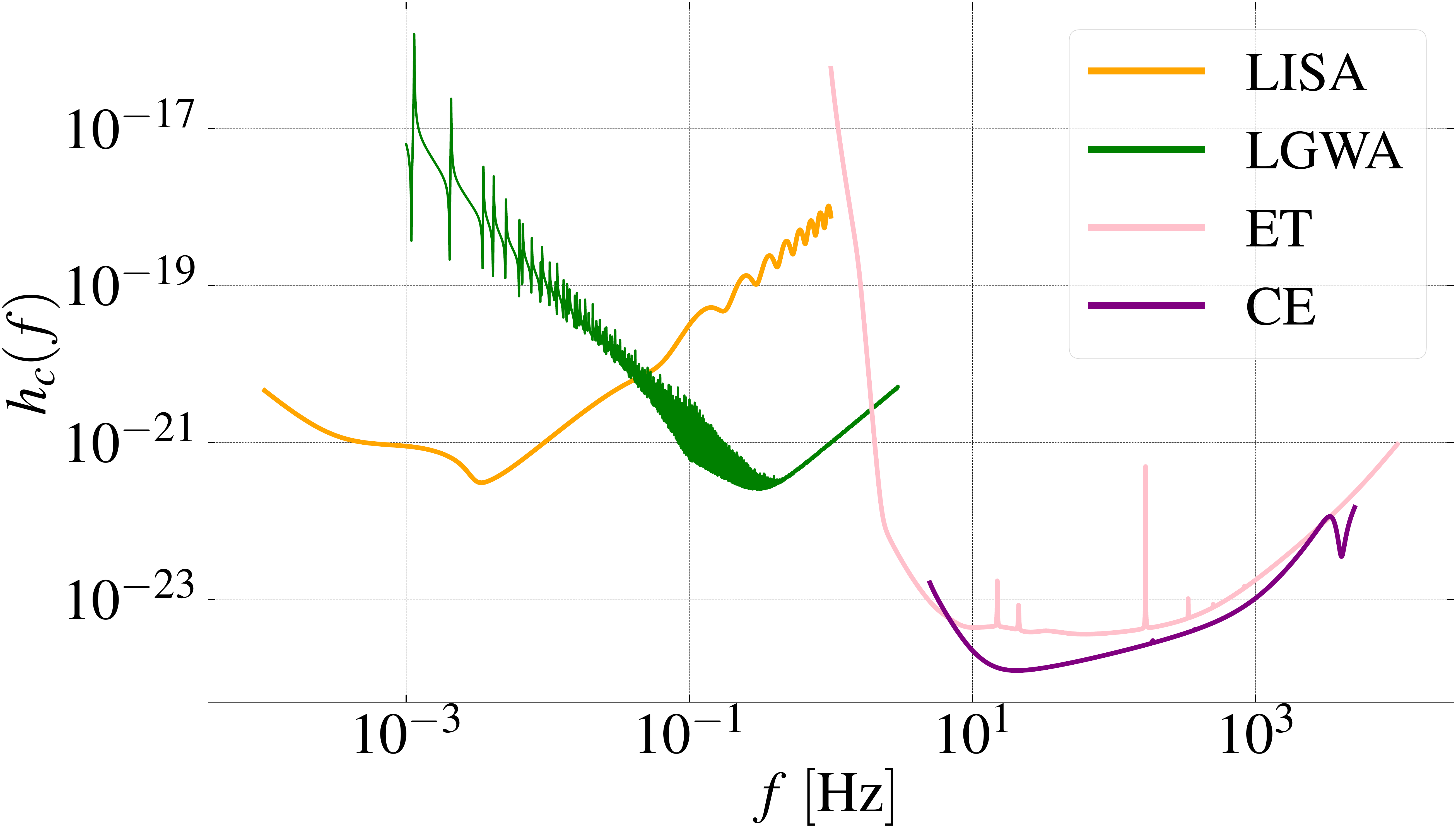}
	\caption{\textit{Left}: Expected horizon distances for the three upcoming detectors (LISA, LGWA and XG) showing their region of detectability in the mass-redshift plane assuming equal mass binaries and averaged over all extrinsic parameters. Dotted (solid) lines show SNR=8 (SNR=20) contours. \textit{Right}: The expected characteristic noise spectra $\sqrt{f  S_n(f)}$, where $S_n(f)$ is the power spectral density of the detectors.}
	\label{fig:multiband}
\end{figure*}

\paragraph{SN and AGN feedback:}

{Accretion onto a central BH need not proceed continuously, because processes such as radiation pressure, star formation, SN feedback, and feedback from the accreting BH itself can regulate or deplete the nuclear gas supply. SN-driven winds can efficiently remove gas from galaxies with shallow potential wells \citep{habouzit_sn_feedback, dubois_sn_feedback}. When the bulge escape velocity is sufficiently low relative to the characteristic wind velocity, SN feedback can deplete the nuclear gas reservoir and suppress BH accretion. This effect is strongest in low-mass galaxies and may delay substantial BH growth until the host becomes sufficiently massive to retain or replenish its gas. More massive galaxies are less susceptible to complete gas removal, although SN feedback can still regulate their gas supply and star-formation histories. The model also accounts for AGN feedback from BHs onto the bulge interstellar medium and the diffuse chemically pristine gas.}

\subsection{The specific models that we consider}
\label{sec:smbh_models_considered}

For our analysis, we consider the following specific models: 

\begin{itemize}

\item \textsc{popIII-d}: This LS model includes delays between halo and galaxy mergers, as well as between galaxy and BH mergers. The latter account for dynamical friction, stellar hardening, interactions involving multiple BHs, and gas-driven migration of bound BH binaries, analogous to planetary migration in gaseous discs \citep{Antonini_2015, popIIId_ref}.

\item \textsc{LS-nod-noSN}: This is an LS model that accounts for the delays between halo and galaxy mergers but does not take into account the delays between galaxy and BH mergers \citep{EPTA:2023xxk, Barausse_2020}. Here ``\textsc{noSN}'' indicates that this model does not take into account the effect of SN feedback on the gas reservoir responsible for the growth of the central BH.   

\item \textsc{LS-nod-SN}: This is an LS model that accounts for the delays between halo and galaxy mergers but does not take into account the delays between galaxy and BH mergers \citep{EPTA:2023xxk, Barausse_2020}. Here ``\textsc{SN}'' indicates that this model takes into account the effect of SN feedback on the gas reservoir responsible for the growth of the central BH.  

\item \textsc{Q3-nod}: This is an HS model that takes into account the delays between halo and galaxy mergers similar to the \textsc{popIII-d} model, but does not assume any delays between galaxy and BH mergers \citep{Antonini_2015, popIIId_ref}. 

\item \textsc{HS-nod-SN-high-accr}: This HS model neglects the delays between galaxy and BH mergers, while retaining those between halo and galaxy mergers, and includes the effect of SN feedback~\citep{EPTA:2023xxk, Barausse_2020}. {Here, the accretion is enhanced by boosting the influx of gas into the nuclear reservoir during star formation events in the spheroid.}

\item \textsc{HS-nod-noSN}: This is an HS model that takes into account the delays between halo and galaxy mergers similar to the \textsc{LS-nod-noSN} model, but does not assume any delays between galaxy and BH mergers. Here ``\textsc{noSN}'' indicates that this model does not take into account the effect of SN feedback~\citep{EPTA:2023xxk, Barausse_2020}. 
\end{itemize}

The expected distribution of source-frame total mass $M$ and cosmological redshift $z$ predicted by these models is shown in Fig.~\ref{fig:merger_rate}. 

\begin{figure*}[tbh]
\centering	
\includegraphics[width=0.65\columnwidth]{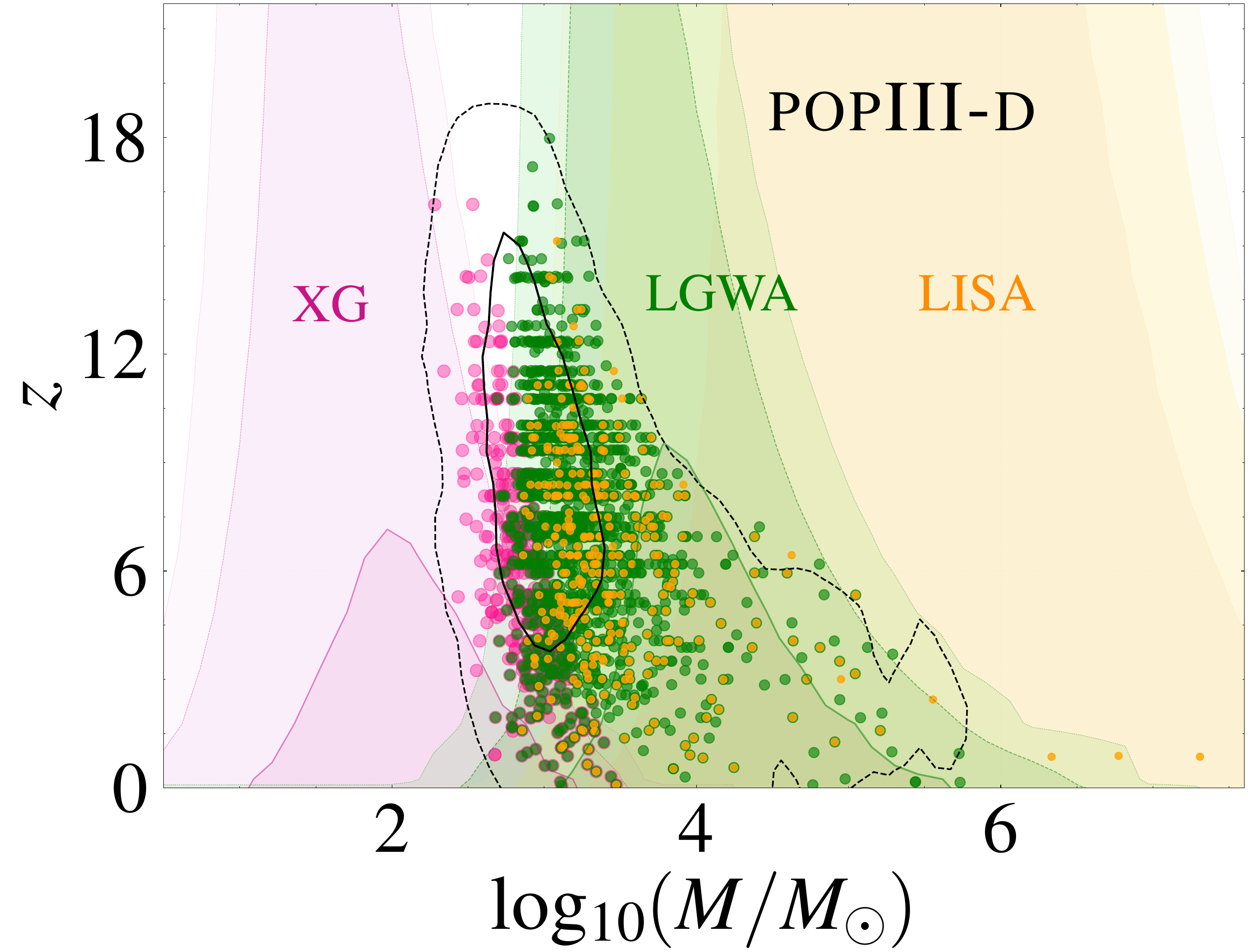}
\includegraphics[width=0.65\columnwidth]{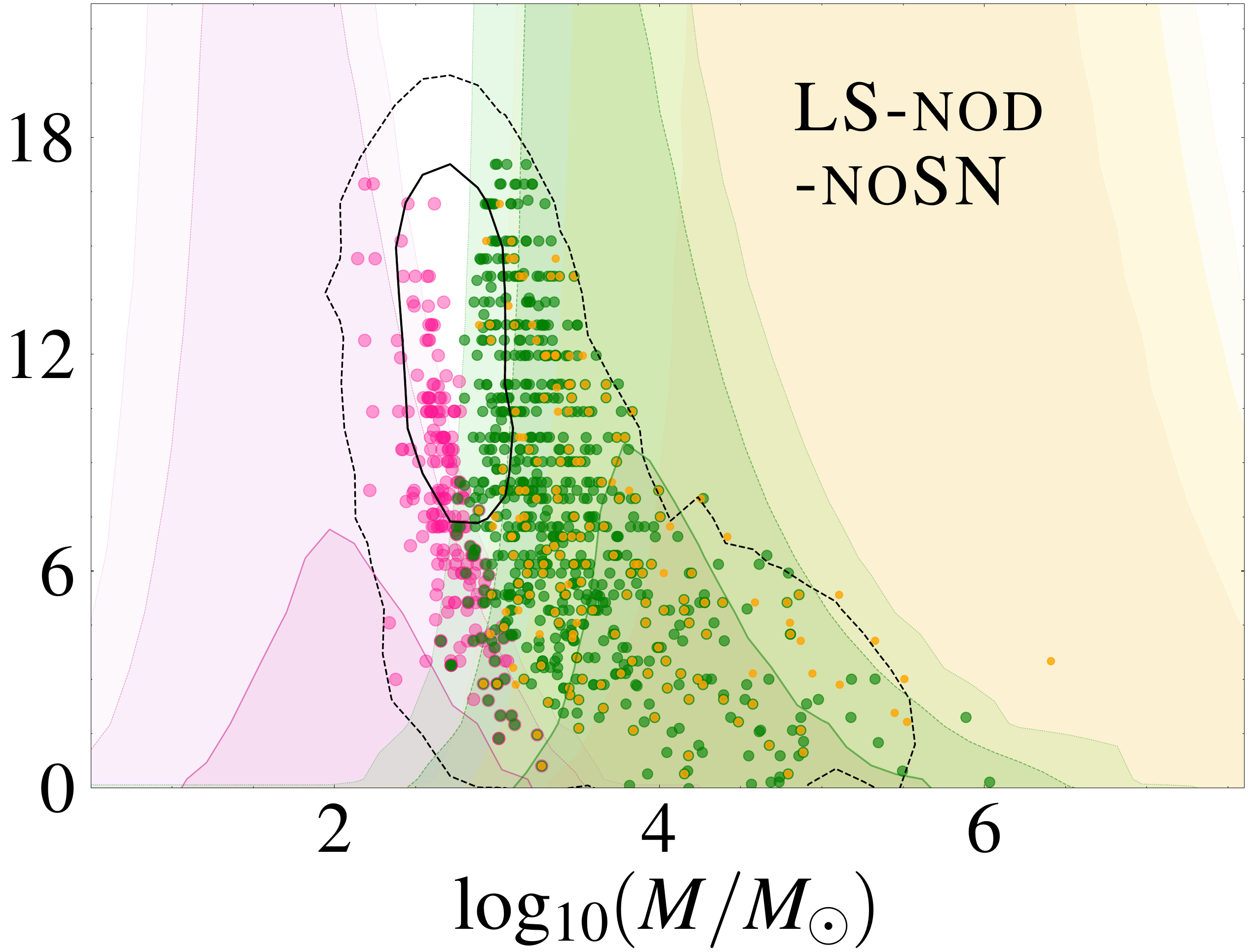}
\includegraphics[width=0.65\columnwidth]{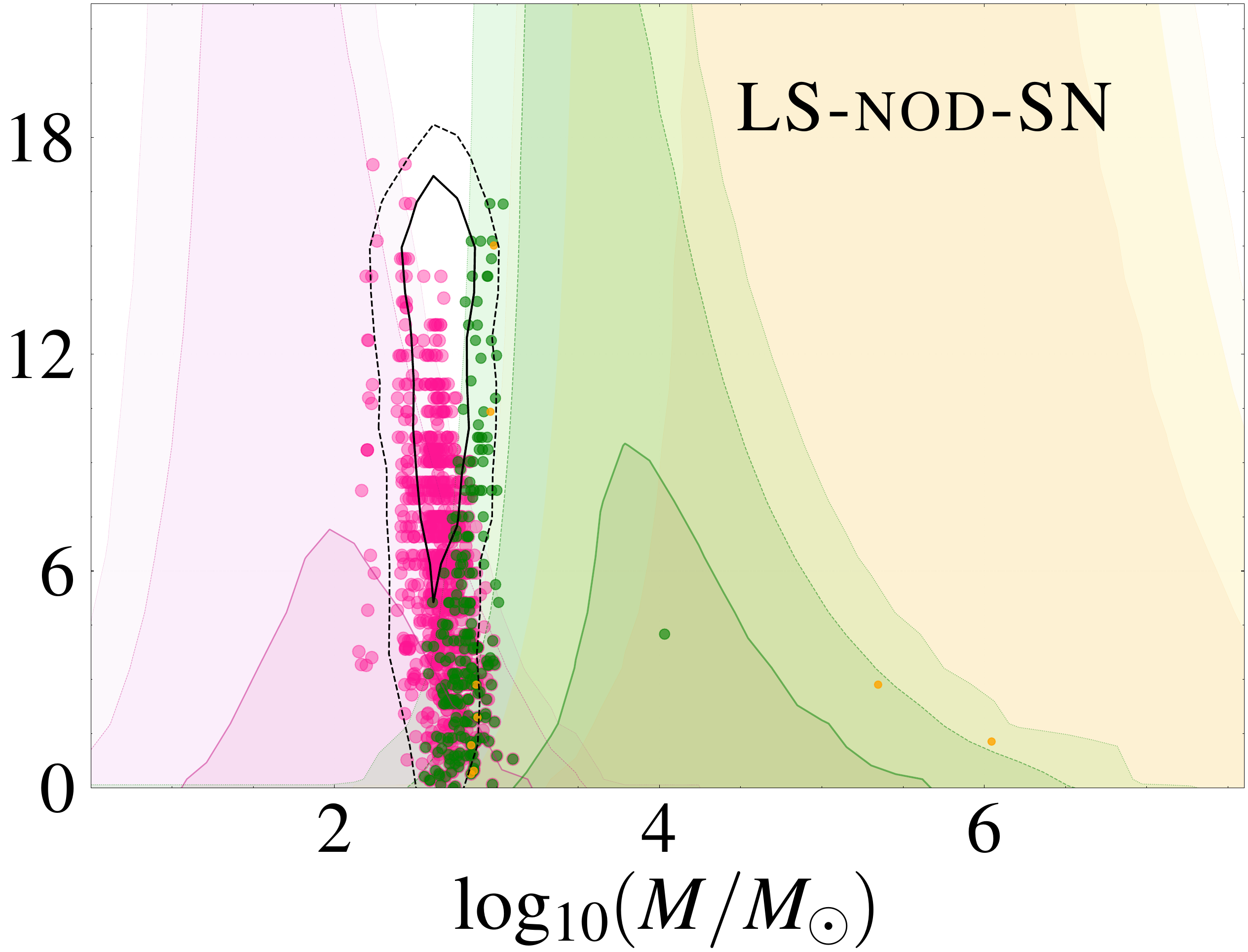}
\includegraphics[width=0.65\columnwidth]{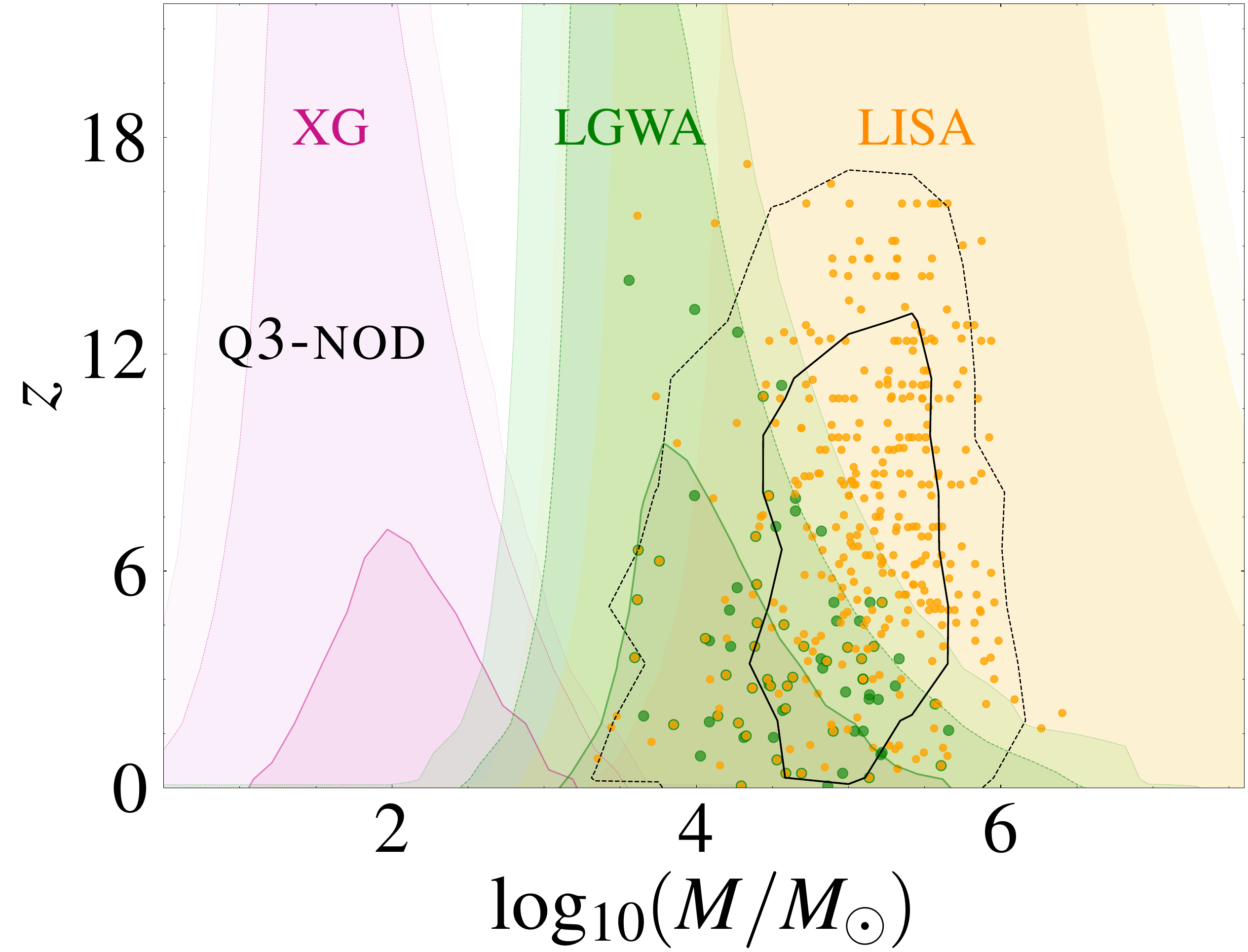}
\includegraphics[width=0.65\columnwidth]{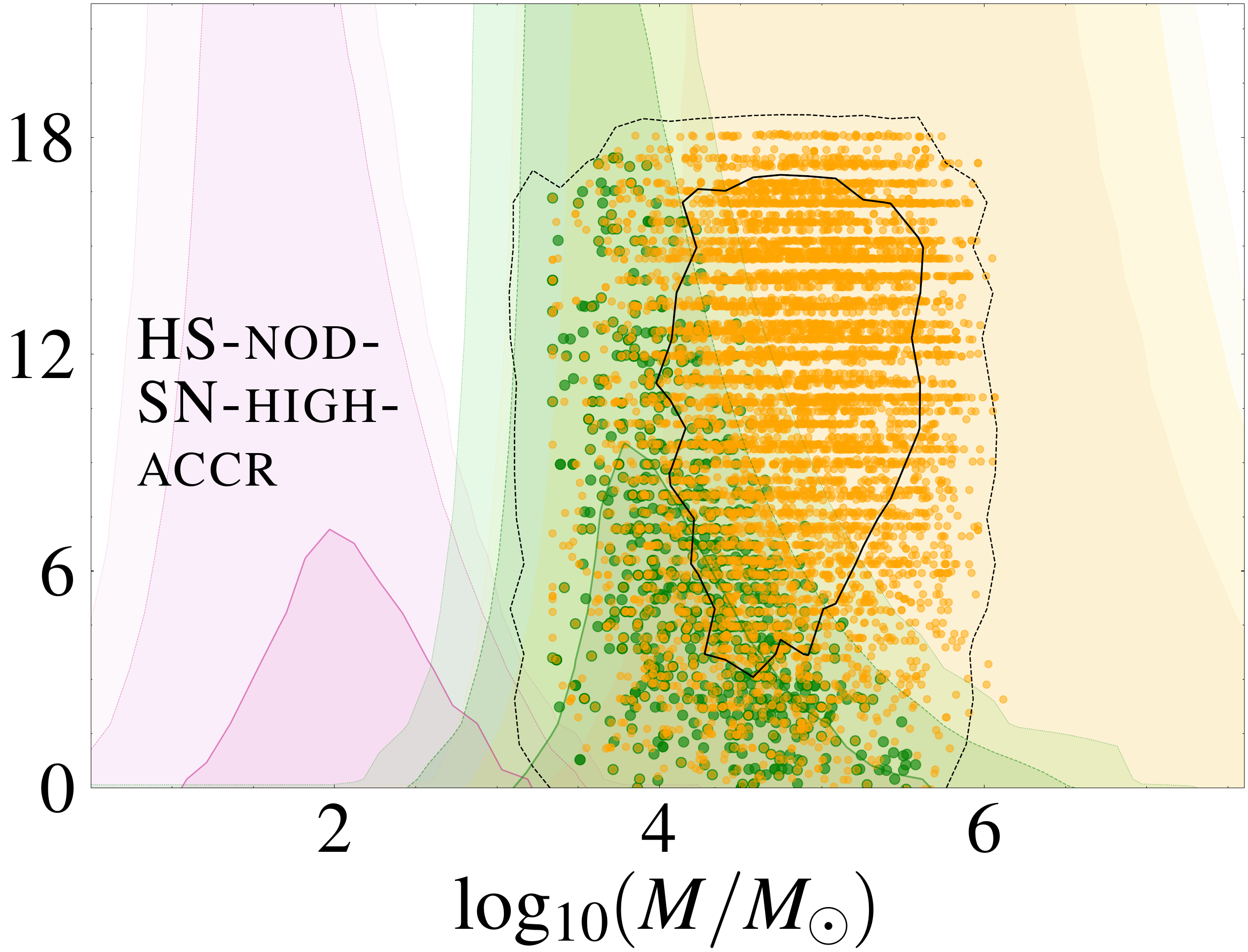}
\includegraphics[width=0.65\columnwidth]{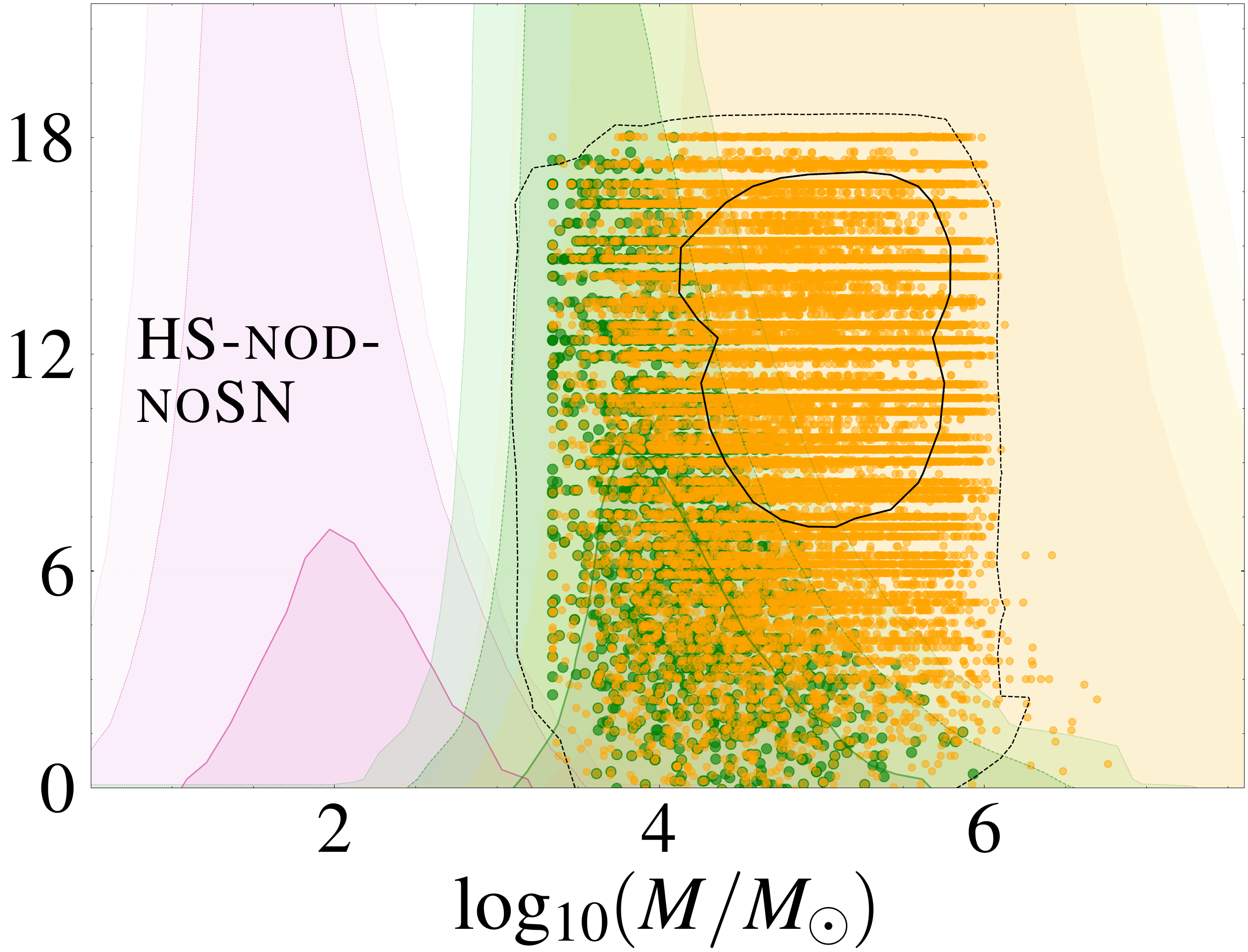}
\caption{BH merger events detectable by individual detectors LISA, LGWA, and XG assuming an SNR threshold of 8. Pink (green) points denote detections by XG (LGWA) over a 10-year observation period, while orange points correspond to detections by LISA over a 4-year mission duration. The black contours represent the underlying true distribution, with the inner (outer) contours indicating the 68$\%$ (99$\%$) percentile for all LS (top) and HS models (bottom).}
\label{fig:ls_det_mod}
\end{figure*}

\section{GW observatories}
\label{sec:detectors}
 
In this work, we consider the following GW observatories. Their expected sensitivity curves are given in Fig.~\ref{fig:multiband}.

\paragraph{Laser Interferometer Space Antenna (LISA):} A space-based laser interferometric observatory scheduled for launch in 2035 operating in the frequency range $\sim 10^{-4} - 0.1$ Hz~\citep{amaroseoane2017laserinterferometerspaceantenna, lisa_universe}. We use the full LISA response function as explained in~\citep{marsat2018fourierdomain} together with SciRDv1 noise, including the unresolved white-dwarf background from Galactic binaries \citep{lisa_unresolved_noise1, lisa_unresolved_noise2}. 
	
\paragraph{Lunar Gravitational Wave Antenna (LGWA):} A proposed first-generation lunar GW detector designed to detect GWs by measuring the vibrations on the Moon induced by passing GWs, operating in the frequency range from $\sim 10^{-3}$ to $\sim{1}$ Hz~\citep{Harms_2021}. Decihertz detectors like LGWA are expected to bridge the observational gap between the millihertz band of space-based detectors such as LISA and the high-frequency band of future terrestrial detectors. The conclusions of this study should also be applicable to other decihertz detectors with similar sensitivities. 
	
\paragraph{Next-generation ground-based detectors (XG):} These involve the Cosmic Explorer (CE)~\citep{CE,evans2021horizonstudycosmicexplorer,hild2011sensitivity} in the USA and the Einstein Telescope (ET)~\citep{ET,maggiore2020etscience} in Europe. They are designed to significantly extend the sensitivity and reach of current ground-based detectors. ET is proposed as an underground detector, optimized to reduce seismic and environmental noise. CE is planned as an overground detector with much longer interferometer arms. Both detectors are expected to operate primarily in the frequency band from a few Hz up to several kilohertz and are supposed to be operational in the late 2030s. In this work, we assume a network of XG detectors comprising one triangular ET configuration with 10-km arms, and two Cosmic Explorer detectors, one with a 20 km arm length and the other with a 40 km arm length. 

{For our main forecasts, we assume nominal observing periods of 10 years for XG and LGWA and 4 years for LISA. We also present results for observing periods of 4 years and 1 year.} 
	
\section{Reconstructing the mass distribution of SMBHs}
\label{sec:reconstruct}

\begin{figure}[tbh]
\centering	
\includegraphics[width=0.95\columnwidth]{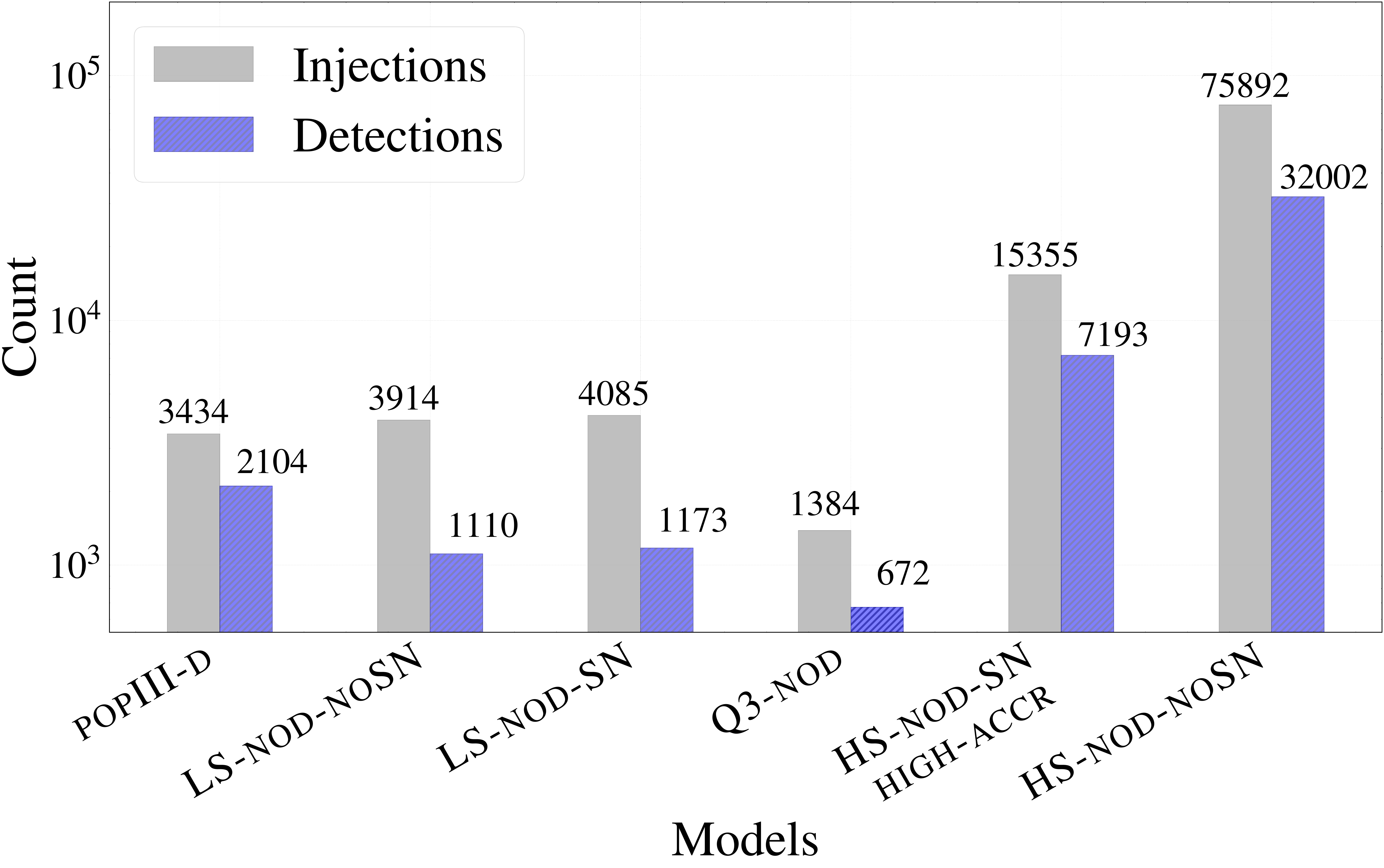}			
\caption{The number of intrinsic and detectable mergers predicted by different models, assuming an observing period of 10 years for XG and LGWA, and 4 years for LISA. Detectable mergers are calculated assuming a network SNR threshold of 8. Numbers shown here are after taking the median over multiple realizations of the injected mergers (and corresponding detections) and after selecting the  $q \geq 0.1$ subpopulation.}
\label{fig:detection_plot}
\end{figure}

\begin{table}[t]
\centering
\begin{tabular}{llcl}
\hline\hline
Parameter & Description & Units & Prior \\
\hline
$\log_{10} M$ & source-frame total mass & $M_\odot$  & $\mathcal{U}(0,\,10)$ \\
$q$                    & mass ratio              & --  & $\mathcal{U}(0.1,\,1)$ \\
$\chi_{1,2}$           & {aligned spins}                   & --  & $\mathcal{U}(-1,\,1)$ \\
$\chi^\perp_{1,2}$           & {in-plane spins}                   & --  & $\mathcal{U}(0,\,1)$ \footnote{We exclude the posterior samples with $(\chi_{1,2})^2 + (\chi^\perp_{1,2})^2 > 1$ as they corespond to unphysical spins. } \\
$z$                  & redshift     & -- & $\mathcal{U}(0,\,30)$ \\ 
$\cos\iota$            & cosine of inclination             & --  & $\mathcal{U}(-1,\,1)$ \\
$\mathrm{RA}$               & right ascension         & rad & $\mathcal{U}(0,\,2\pi)$ \\
$\sin\mathrm{DEC}$           & sine of declination             & --  & $\mathcal{U}(-1,\,1)$ \\
$\psi$                 & polarization angle      & rad & $\mathcal{U}(0,\,\pi)$ \\
$\phi_c$               & coalescence phase       & rad & $\mathcal{U}(0,\,2\pi)$ \\
$t_{c}$                  & coalescence time        & yrs  & $\mathcal{U}(0,\,10)$ \\
\hline\hline
\end{tabular}
\caption{Priors on the source parameters. $\mathcal{U}(a,b)$ denotes a uniform distribution in the interval $(a,b)$. We convert $t_c$ and its prior bounds from years to the corresponding GPS time units.}
\label{tab:priors}
\end{table}

\begin{figure*}[tbh]
\centering	
\includegraphics[width=2.1\columnwidth]{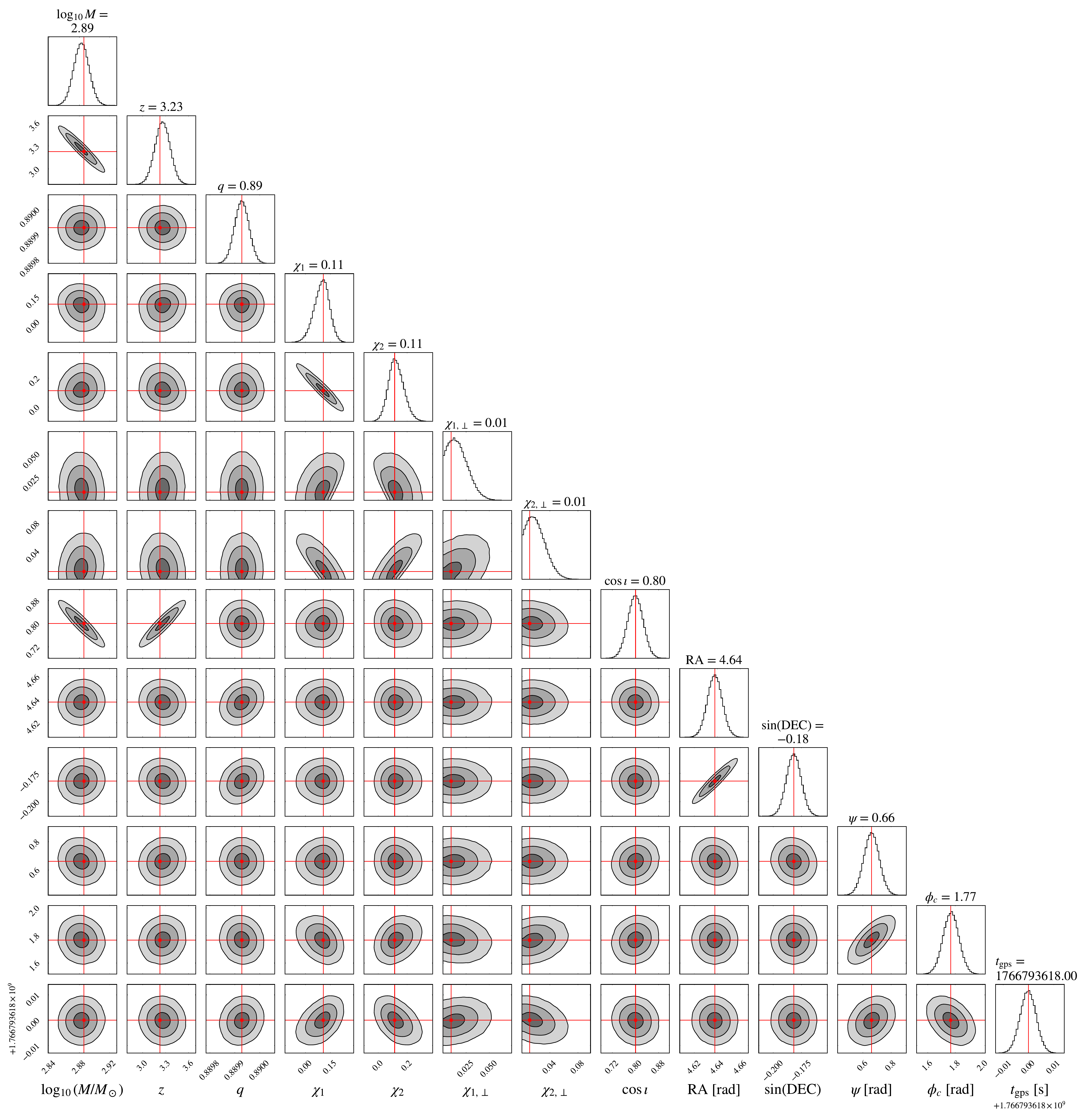}
\caption{Corner plot of the posterior distributions of the binary parameters ($\log_{10} M, z, q, \chi_1, \chi_2, \chi^\perp_1, \chi^\perp_2, \cos \iota, \mathrm{RA}, \sin \mathrm{DEC}, \psi, \phi, $ and $t_{\rm gps}$) obtained from one simulated BBH merger event. {We also include the weak-lensing bias in the $d_L$ measurement. The red lines denote the true values of the binary parameters. The posteriors include the systematic bias introduced by weak lensing, and are not marginalized over the weak-lensing errors.} }
\label{fig:full_corner}
\end{figure*}

\begin{figure}[t]
\includegraphics[width=1.025\columnwidth]{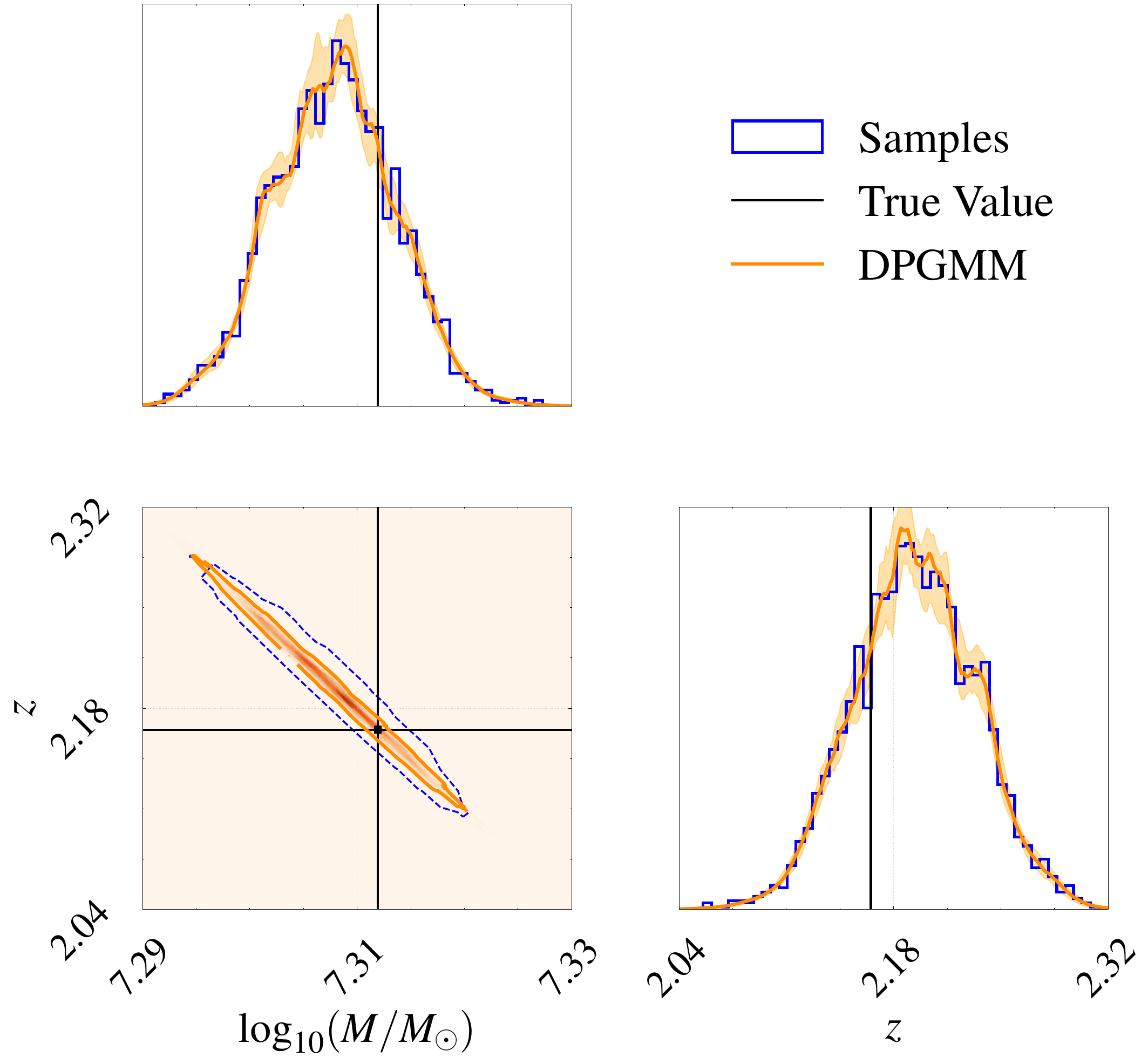}
\caption{Reconstructed posterior distribution of the source-frame mass and redshift (marginalized over all other parameters) from one simulated BBH event. Black lines show the true values, blue lines correspond to the marginalized 1D/2D distributions (90\% credible regions) obtained using a histogram of the posterior samples, while the orange lines correspond to the  DPGMM reconstruction and error bars (90\% error). }
\label{fig:2d_corner_post_recon}
\end{figure}

\begin{figure}[t]
	\centering	
	\includegraphics[width=1\columnwidth]{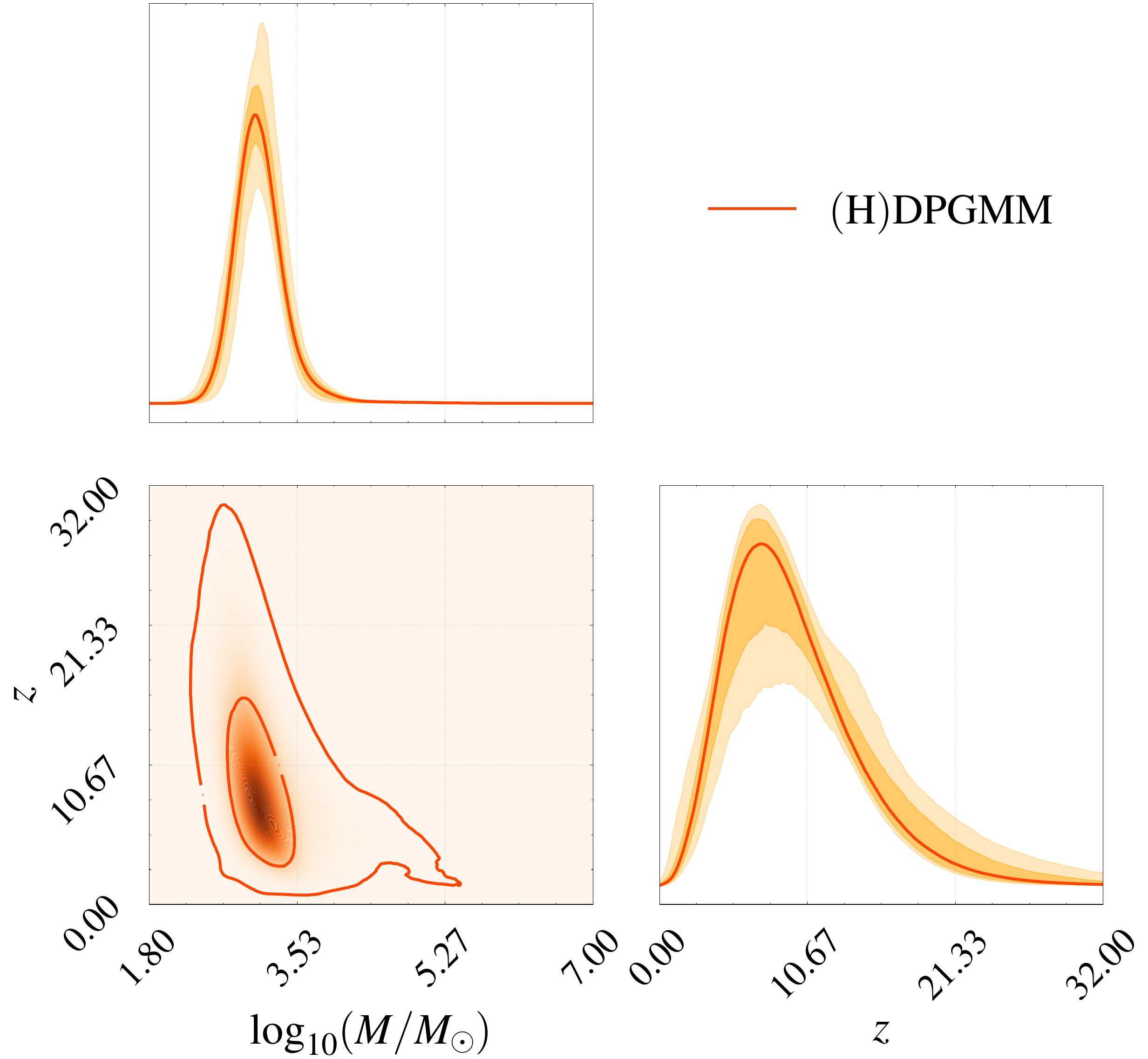}
	\caption{HDPGMM reconstruction of the true distribution for the popIII-d model. The 2D plot shows the $68 \%$ and $99 \%$ contours of the reconstructed true distribution. In the marginalized 1D plots, the red line is the median reconstruction, while the shaded regions are the $68 \%$ and $95 \%$ credible regions.}
	\label{fig:hdpgmm_recon_popIII}
\end{figure}

\begin{figure*}[tbh]
\centering	
\includegraphics[width=0.7\columnwidth]{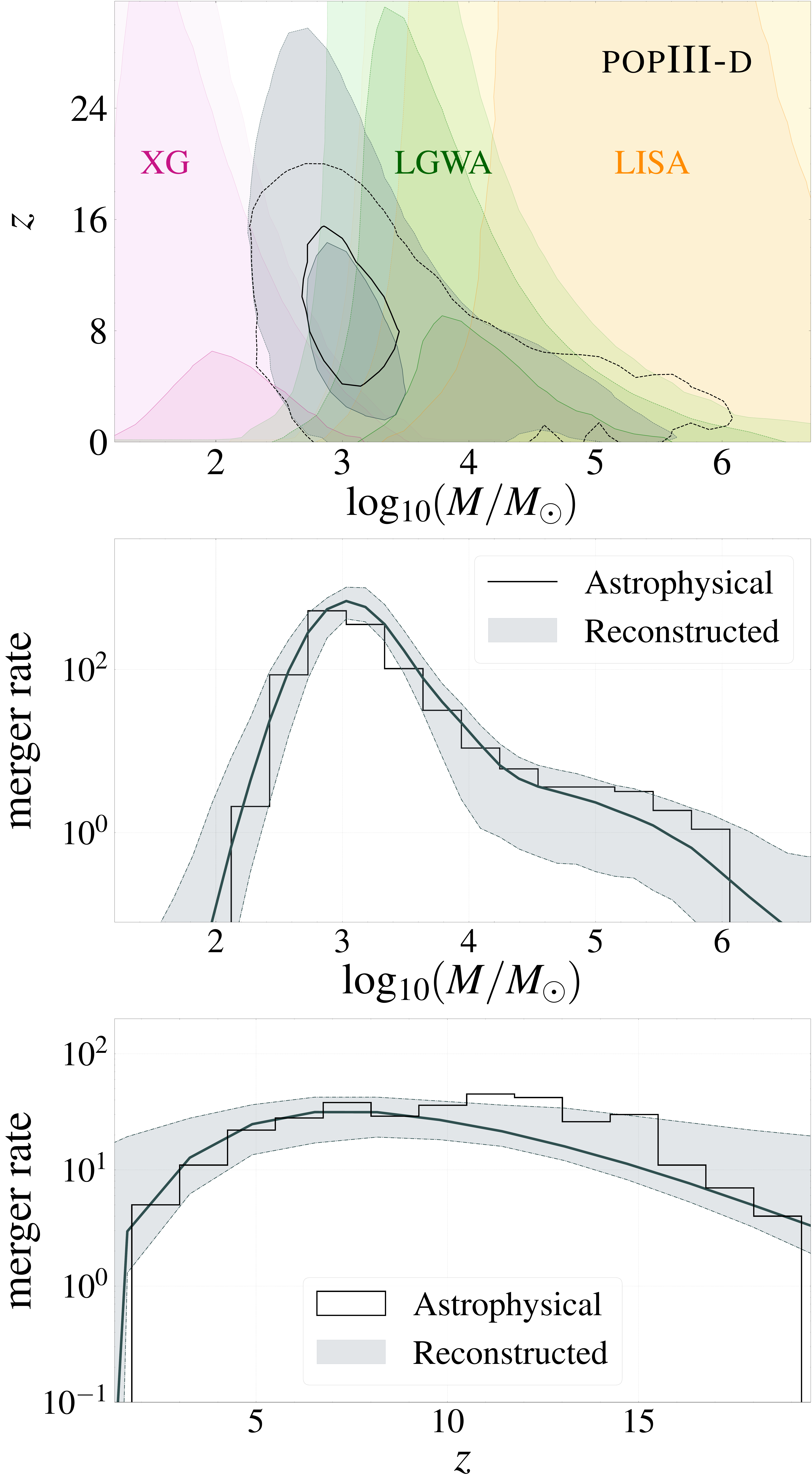}
\includegraphics[width=0.701\columnwidth]{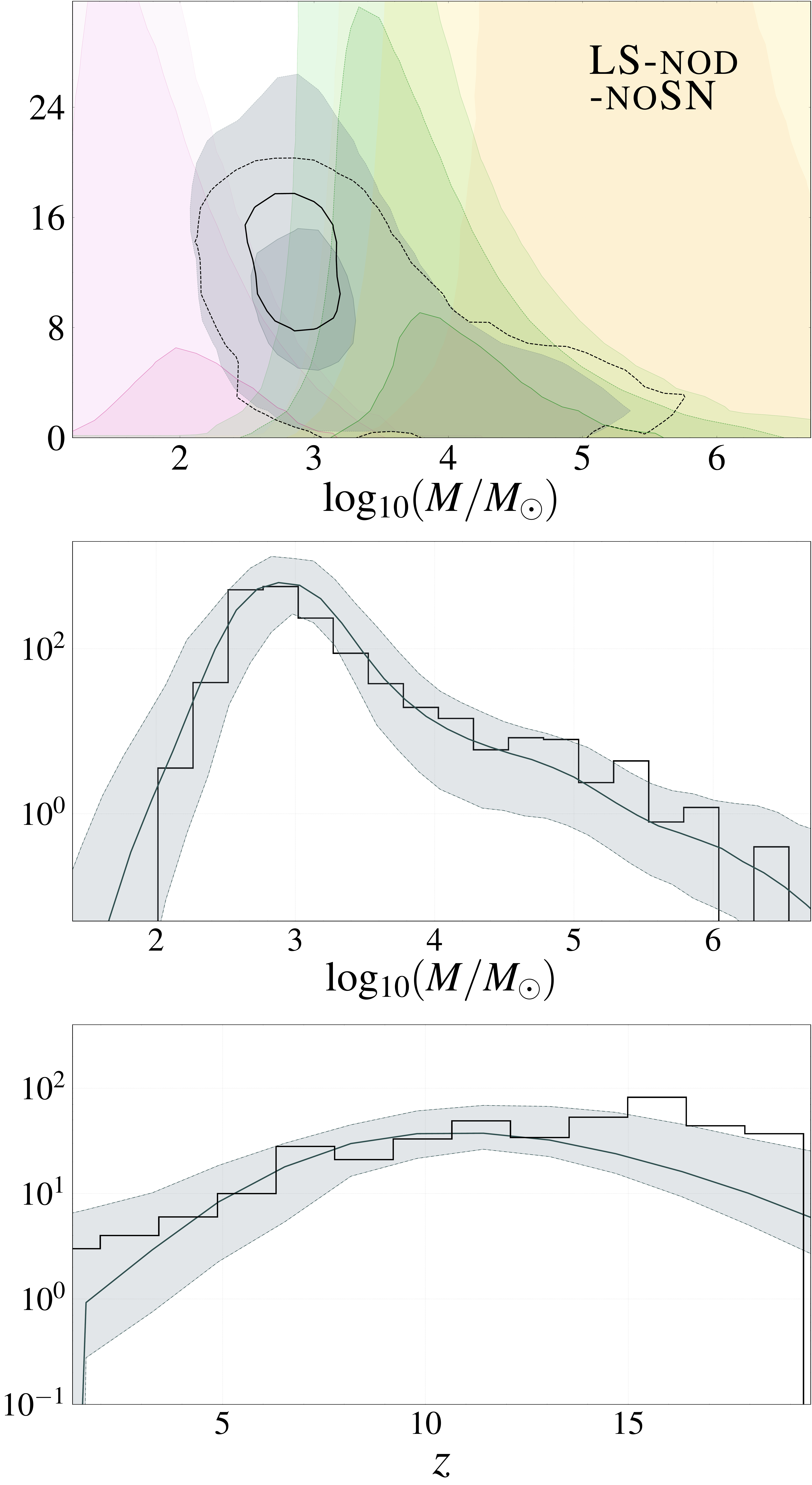}
\includegraphics[width=0.7\columnwidth]{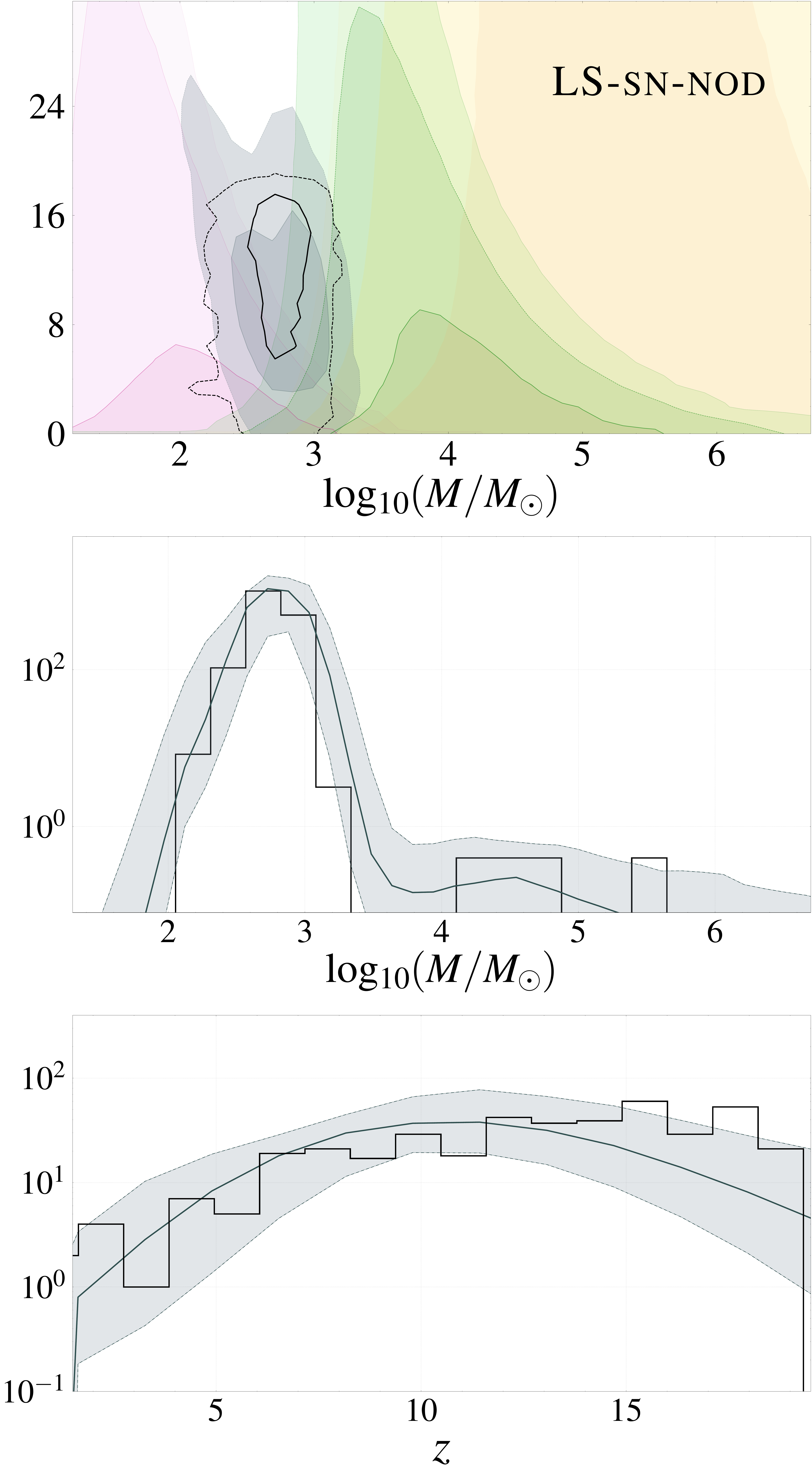}
\caption{\textit{Top row}: The observer-frame merger-rate density  ${d^3N}/[{dt_{\rm o}\,d\log_{10}M\,dz}]$  reconstructed from simulated multiband GW observations spanning a period of 10 yrs (with LISA operating for 4 years). The black contours show the actual simulated LS population (black text label shows the model name) and the grey regions show the reconstruction ($68\%$ and $99\%$ credible regions). Colored contours show the selection function (contours of $p_\mathrm{det} = 0.9$, 0.5 and 0.1) of the XG, LGWA and LISA detectors. \textit{Middle and bottom rows}: One-dimensional reconstructed rate densities as a function of the total mass $M$ (middle) and redshift $z$ (bottom), obtained by marginalizing the 2D distribution.  Shaded gray regions denote $95\%$ credible intervals, and black histograms correspond to the true astrophysical population, while the smooth black curves correspond to median reconstructions.}
\label{fig:ls_recons_plots}
\end{figure*}

\begin{figure*}[tbh]
\centering	
\includegraphics[width=0.7\columnwidth]{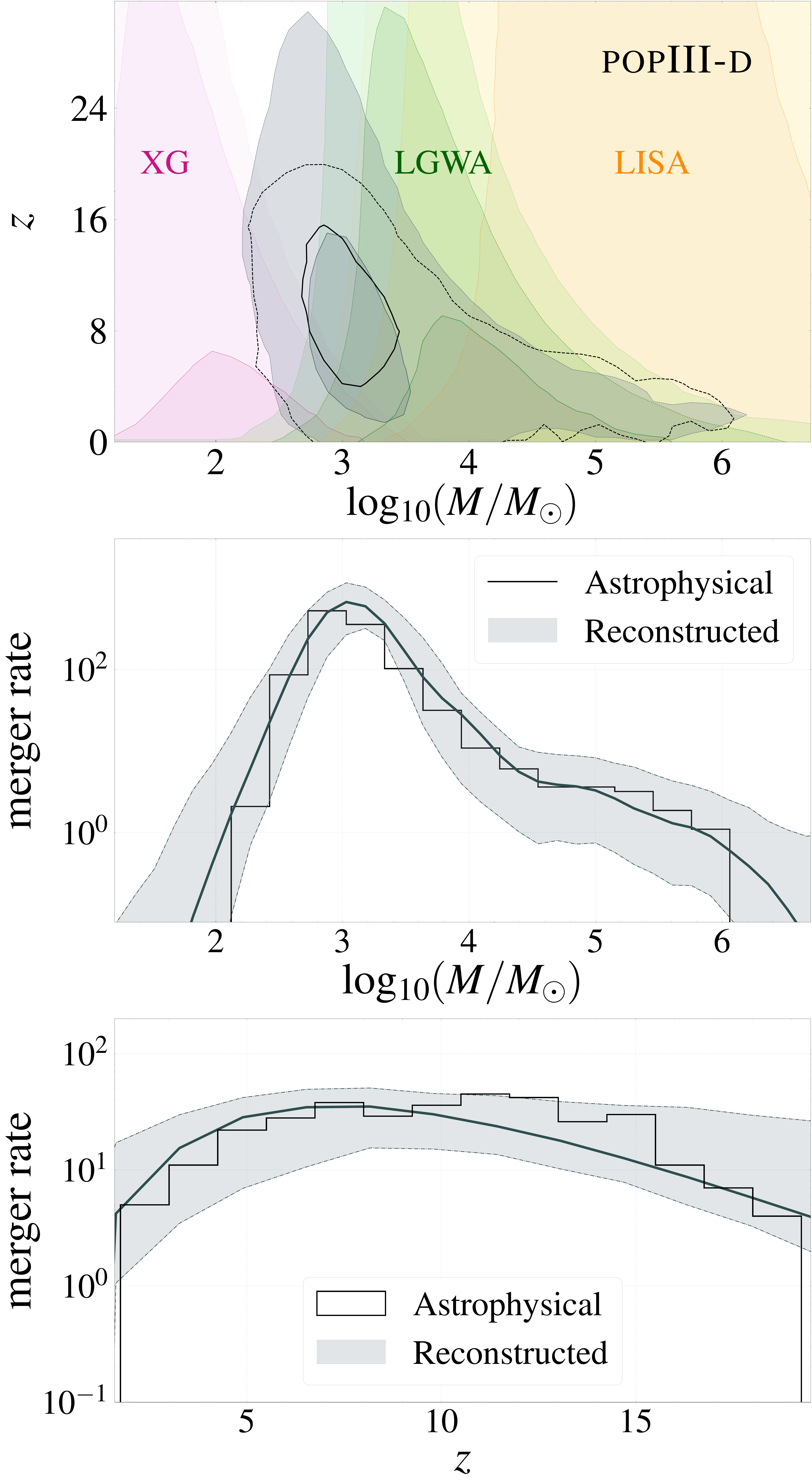}
\includegraphics[width=0.701\columnwidth]{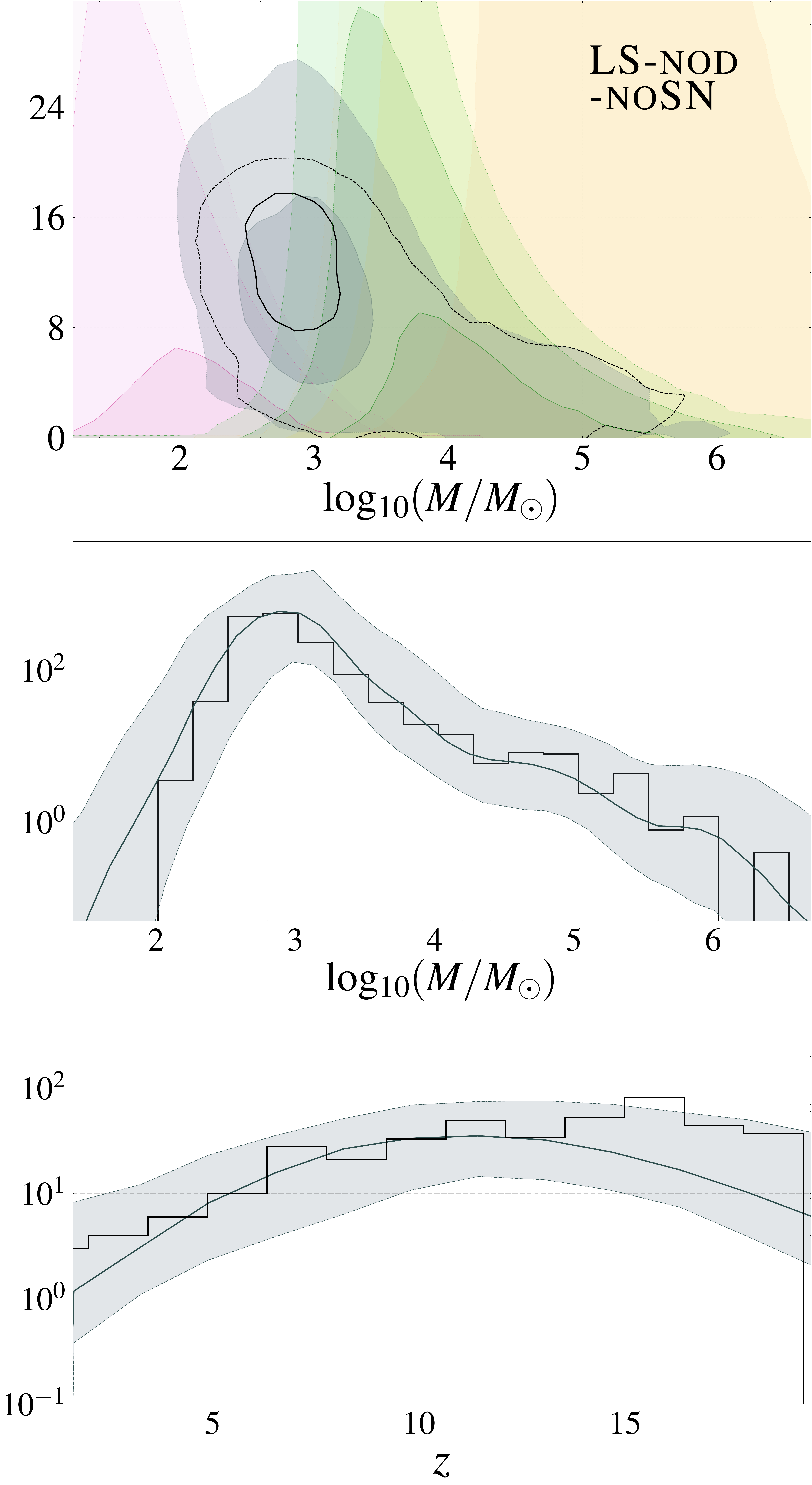}
\includegraphics[width=0.7\columnwidth]{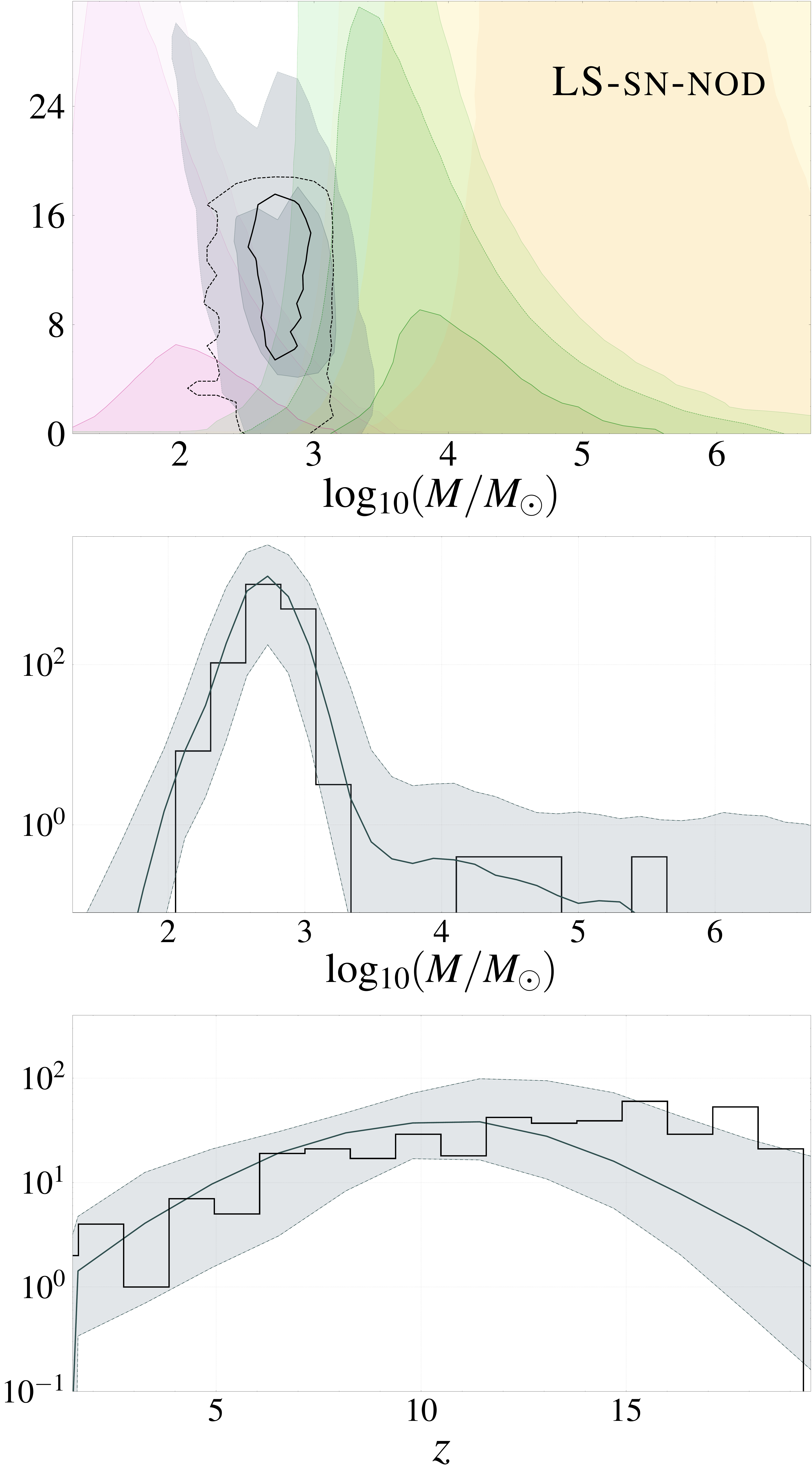}
\caption{Same as Fig.~\ref{fig:ls_recons_plots}, except that the plots correspond to simulated multiband GW observations spanning a period of 4 yrs.}
\label{fig:ls_recons_plots_4yrs}
\end{figure*}

\begin{figure*}[tbh]
\centering	
\includegraphics[width=0.7\columnwidth]{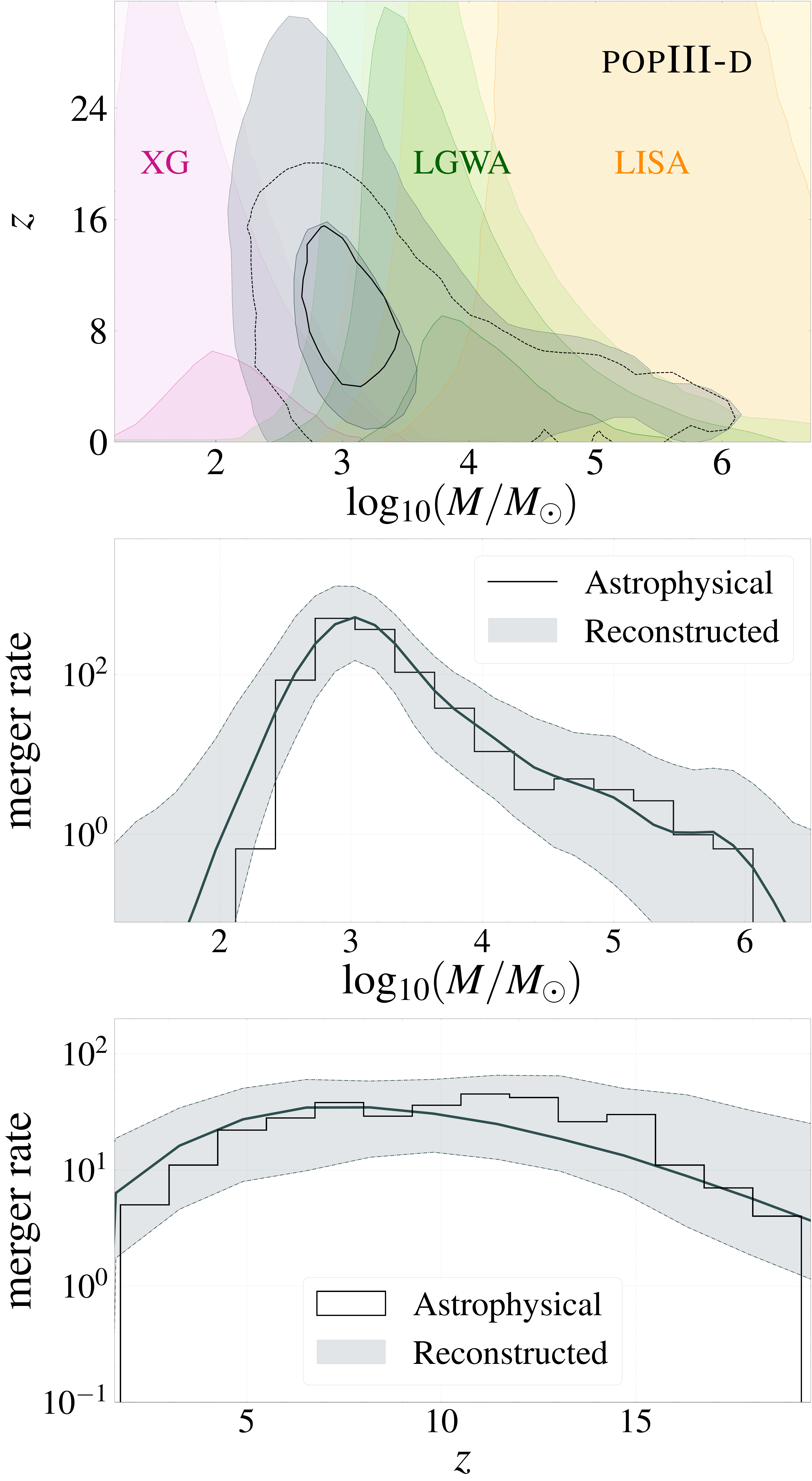}
\includegraphics[width=0.701\columnwidth]{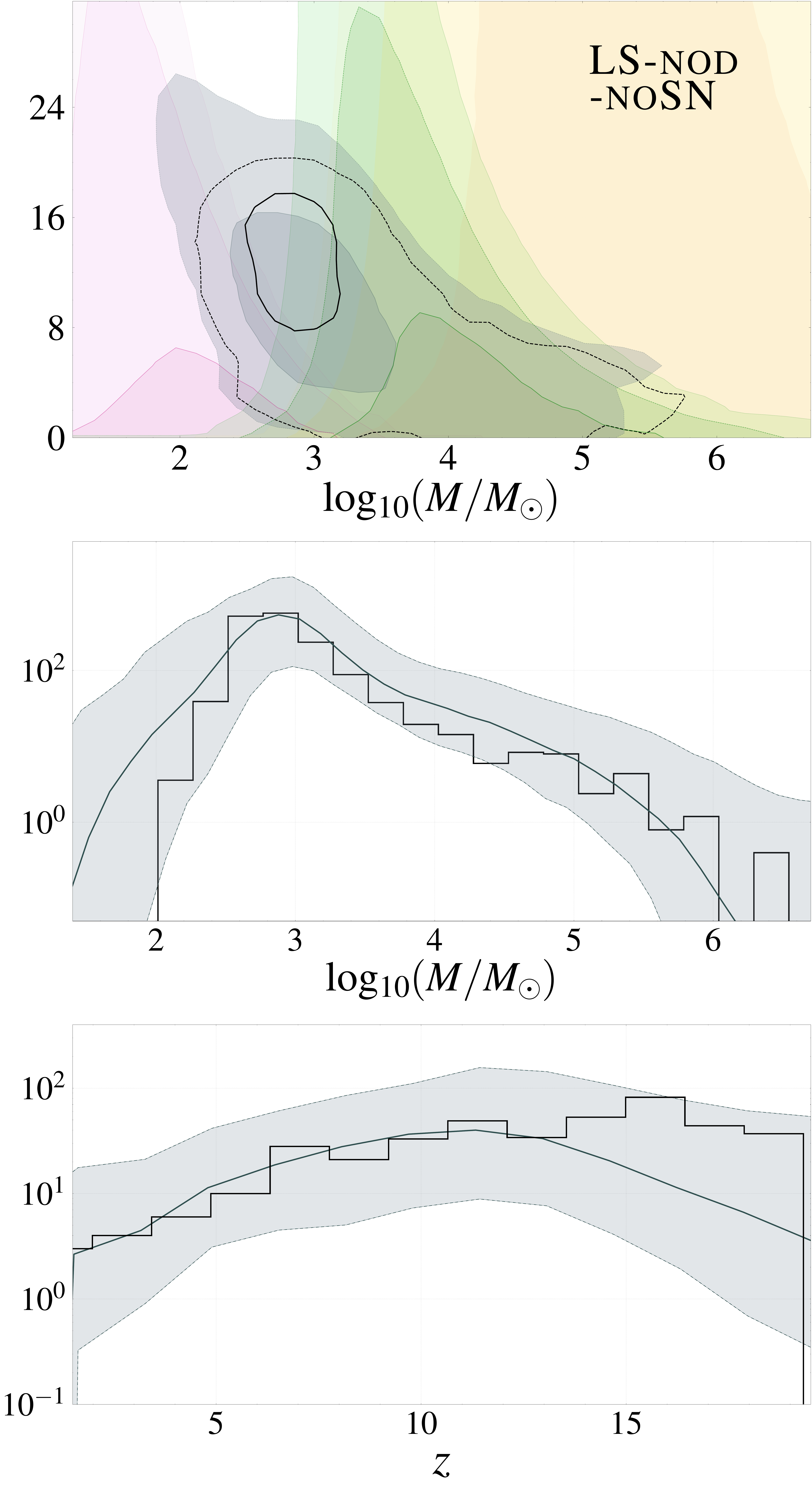}
\includegraphics[width=0.7\columnwidth]{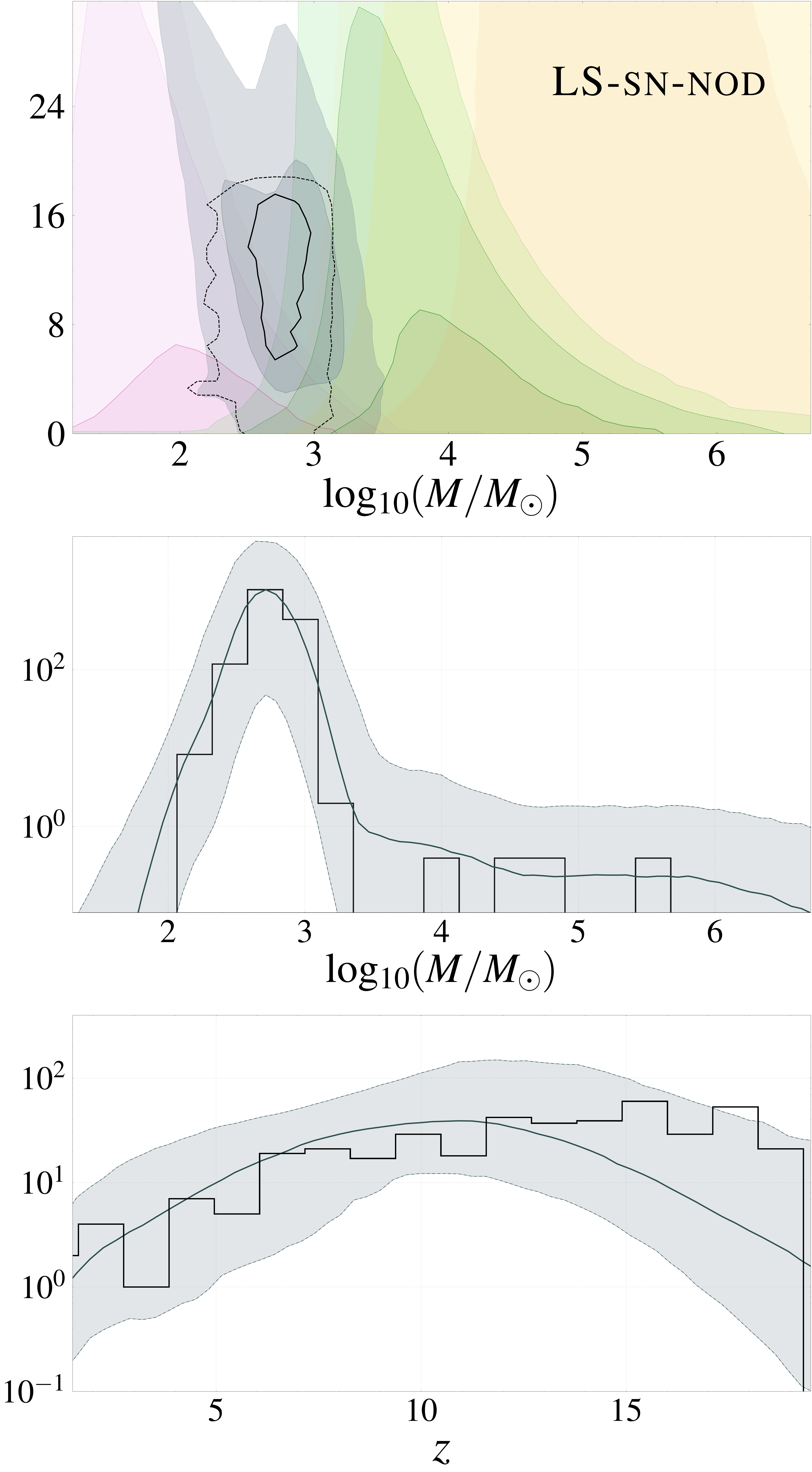}
\caption{Same as Fig.~\ref{fig:ls_recons_plots}, except that the plots correspond to simulated multiband GW observations spanning a period of {1 yr}.}
\label{fig:ls_recons_plots_1yrs}
\end{figure*}

\begin{figure*}[tbh]
\centering	
\includegraphics[width=0.7\columnwidth]{plots/q3nod_full_reconst_plot.pdf}
\includegraphics[width=0.7\columnwidth]{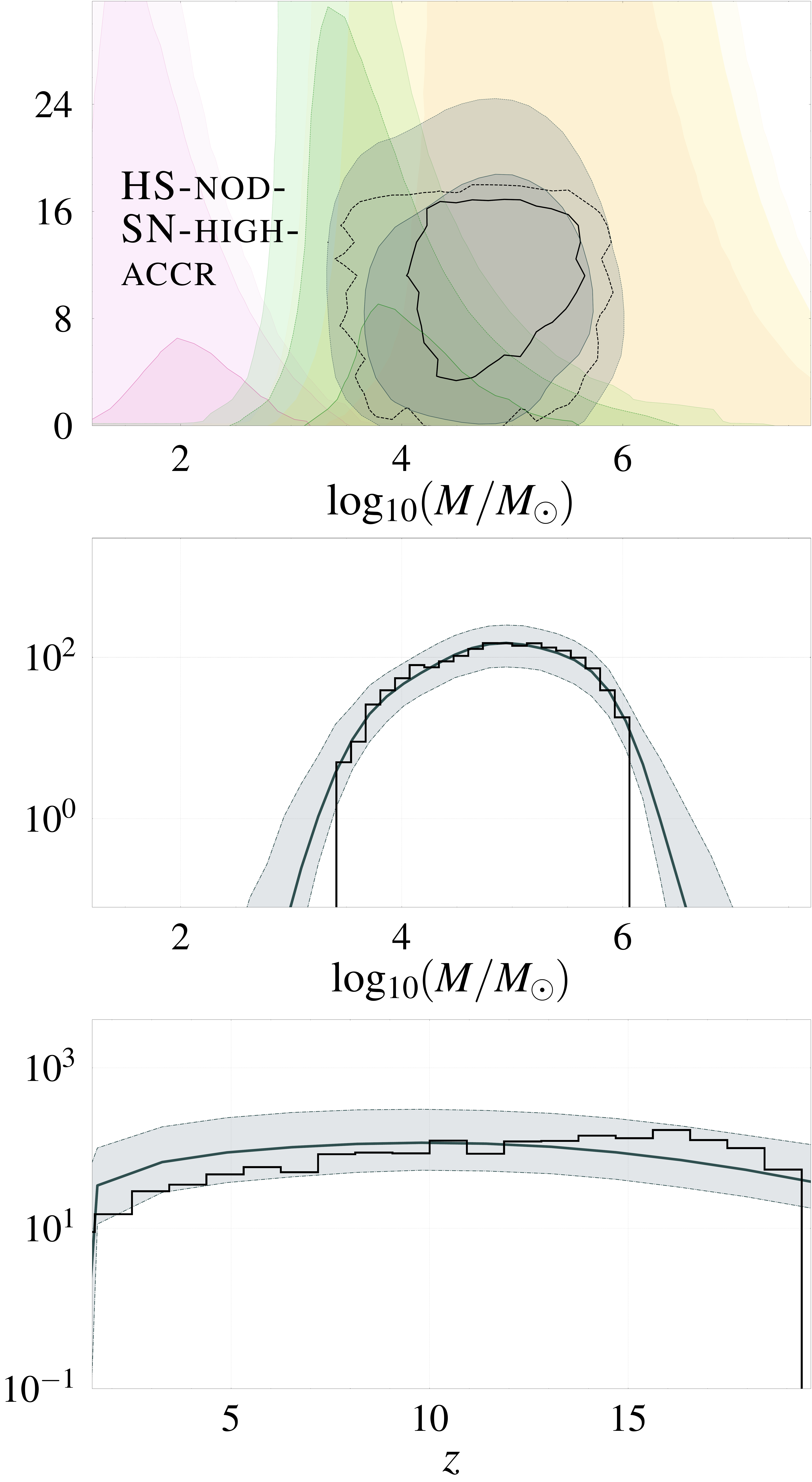}
\includegraphics[width=0.7\columnwidth]{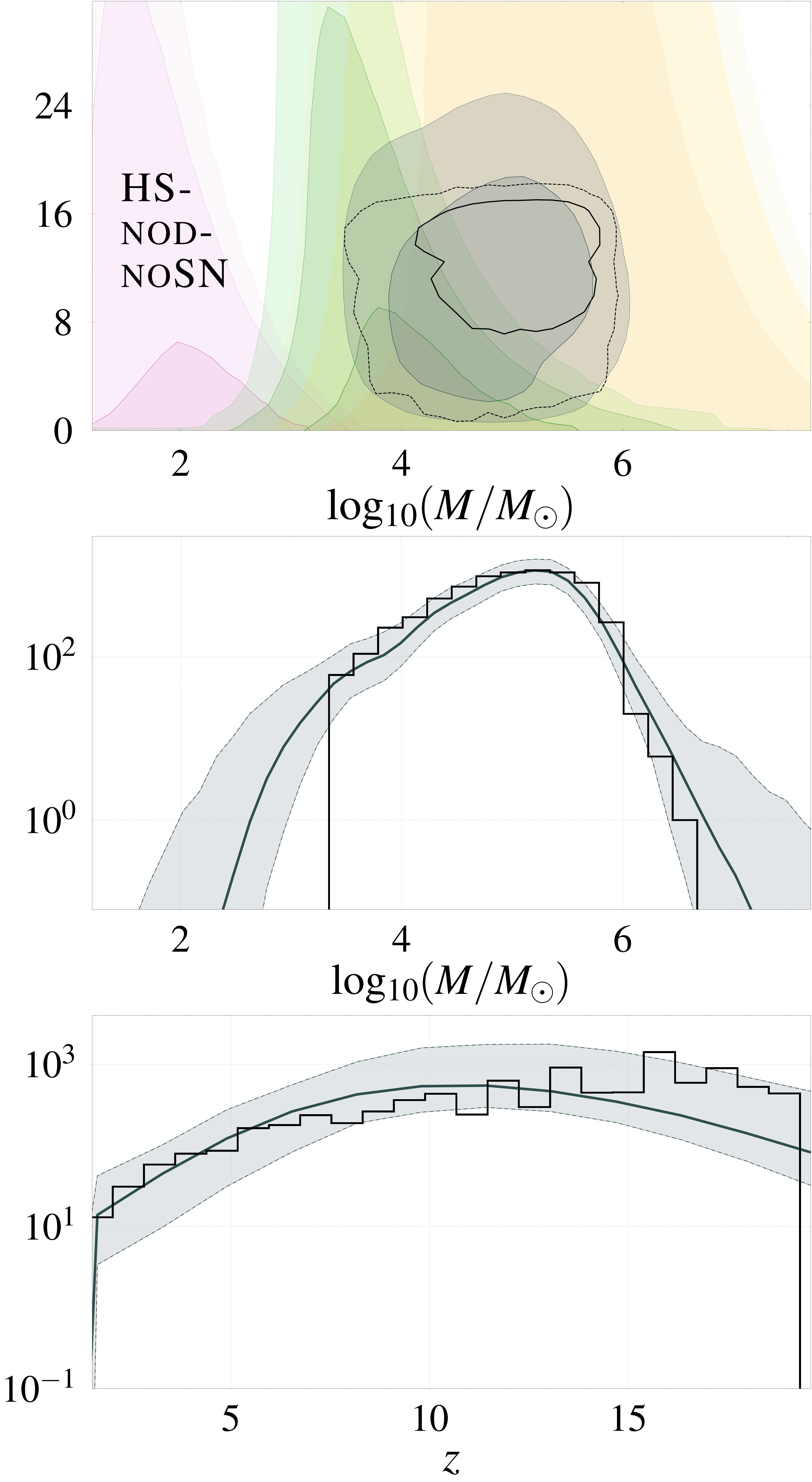}
\caption{Same as Fig.~\ref{fig:ls_recons_plots}, except that the plots correspond to the HS models (assuming an observation period of 10 years).}
\label{fig:hs_recons_plots}
\end{figure*}

\begin{figure*}[tbh]
\centering	
\includegraphics[width=0.7\columnwidth]{plots/q3nod_full_reconst_plot_1yrs.pdf}
\includegraphics[width=0.7\columnwidth]{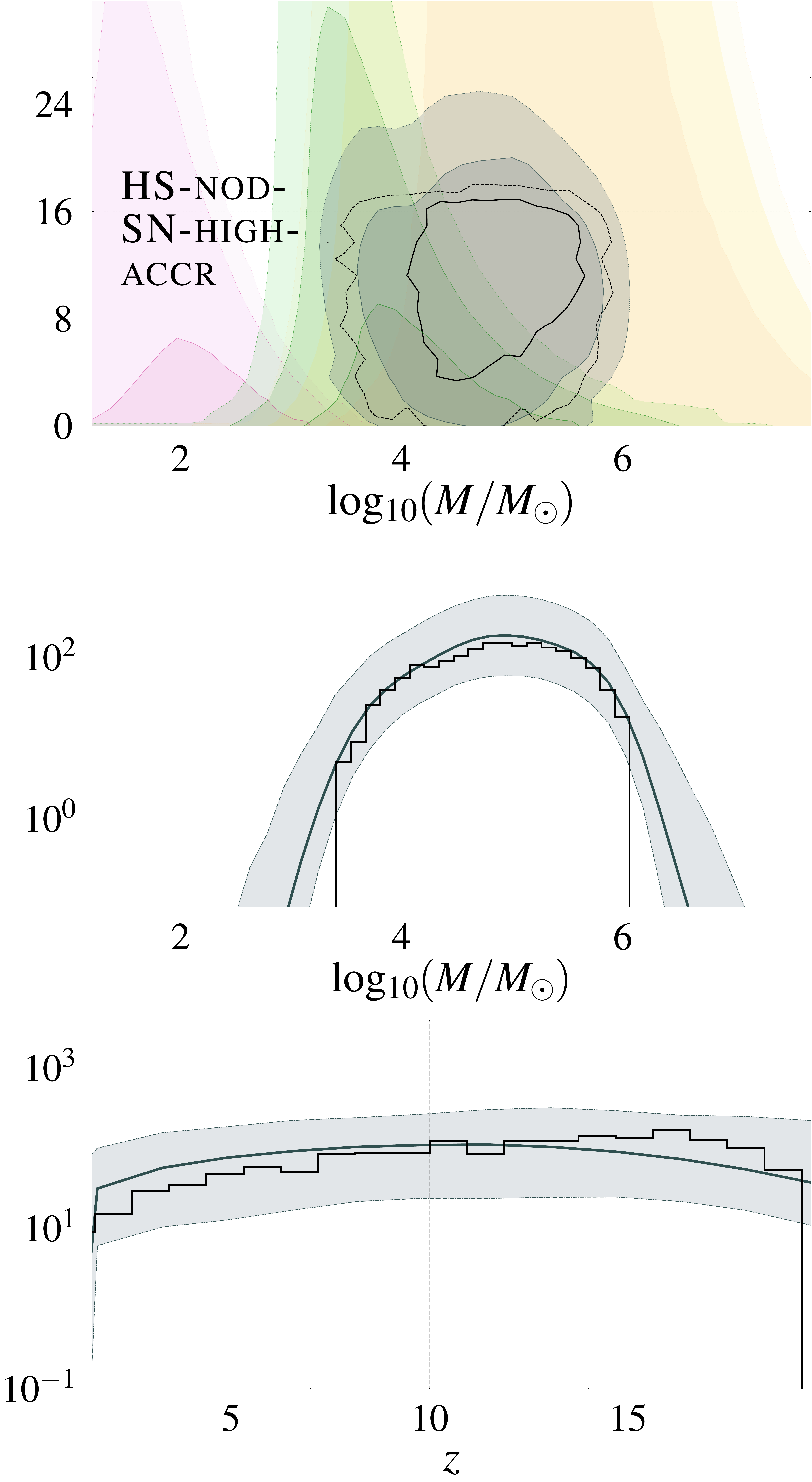}
\includegraphics[width=0.7\columnwidth]{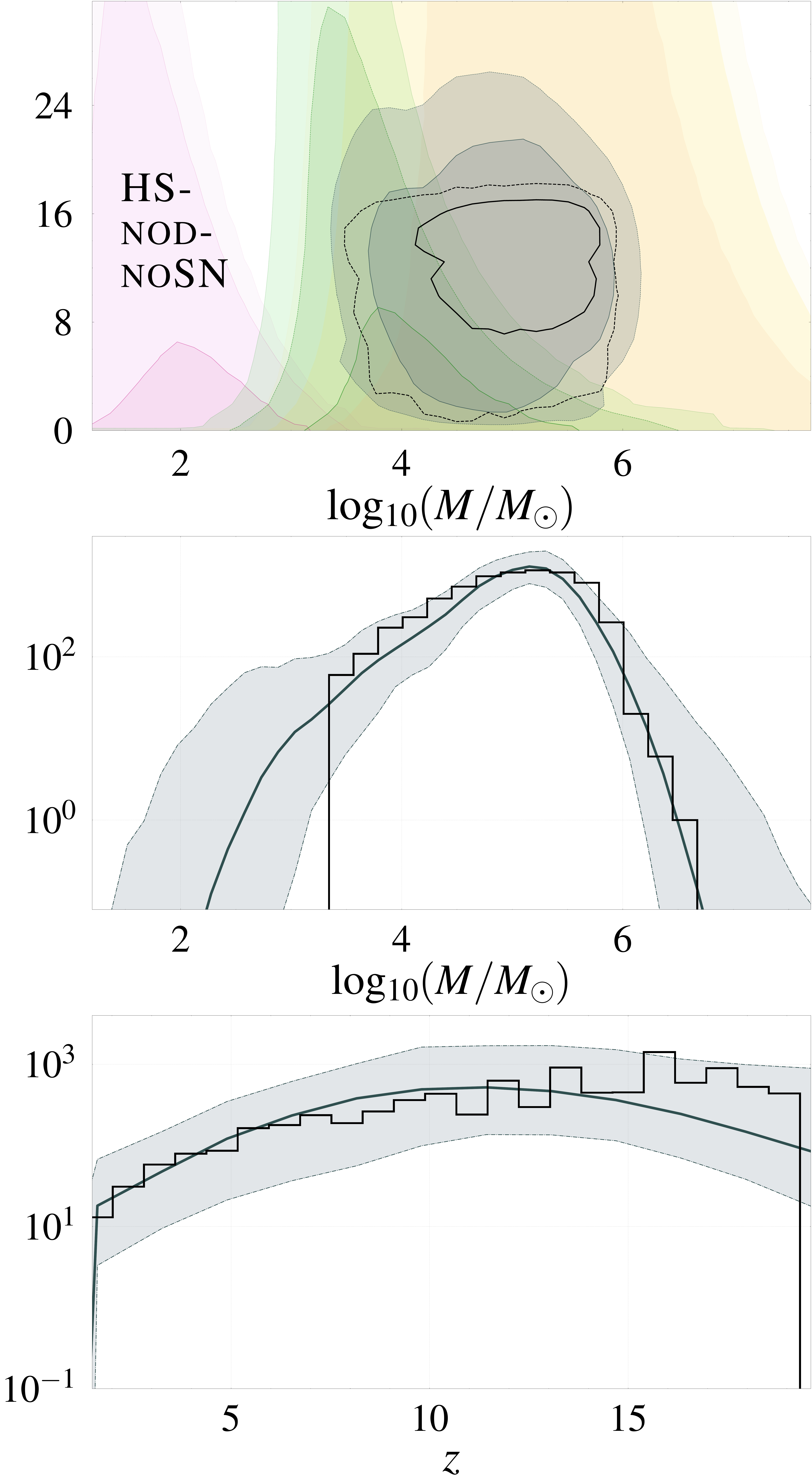}
\caption{Same as Fig.~\ref{fig:ls_recons_plots_1yrs}, except that the plots correspond to the HS models (assuming an observation period of 1 year).}
\label{fig:hs_recons_plots_1yr}
\end{figure*}

The astrophysical models described in Sec.~\ref{sec:smbh_models_considered} provide the expected BH merger-rate density $d^2N/(dV_c\,dt_s)$; i.e., the number of mergers per unit comoving volume per unit source-frame time. We draw $N_\mathrm{sim}$ random 10-yr realizations of BH mergers from the astrophysical model, i.e., assuming a total observation time $dt_o = 10$ yrs. XG and LGWA observe all the events in each realization, while LISA, with a mission duration of 4 yrs, only observes the events merging within the first 4 yrs. Note that a time interval $dt_o$ in the observer's frame is related to a time interval $dt_s$ in the source frame by $dt_o = (1+z) \, dt_s$ due to cosmological time dilation. 

The astrophysical models provide the source-frame total mass $M \equiv m_1+m_2$, mass ratio $q \equiv m_2/m_1 \leq 1$ (with $m_1 \geq m_2$), spin magnitudes $a_1, a_2$, spin tilt angles $\theta_1, \theta_2$ and redshift $z$. We parametrize the spins in terms of their components parallel to the orbital angular momentum, i.e., $\chi_i \equiv a_i \cos \theta_i$, and components perpendicular to the orbital angular momentum, i.e.,  $\chi^{\perp}_{i} \equiv a_i \sin \theta_i$, where $i=1,2$. The rest of the parameters, such as the sky location and orientation of the binary, time of arrival of the signal at the detector, etc., are randomized. 

{Due to the limitations of the waveform approximants used in this study~\citep{IMRPhenomHM_London_2018, IMRPhenomXPHM_Pratten_2021}, we also consider only binaries with mass ratios $q \geq 0.1$. The fraction of binaries with $q < 0.1$ is negligible in the LS models that we consider ($\sim 1-5\%$). For the HS models this fraction ($\sim 20-25\%$) is not negligible. Thus, for the HS models, both the ``true'' and the reconstructed distributions refer to the $q\geq0.1$ subpopulation only. More details are given in Appendix~\ref{sec:sim_mergers}.}

We then compute the optimal SNR of detecting these mergers in various detectors. We treat these detectors as part of a network, with some overlap in the frequency band. The network SNR $\rho$ of an event is given by 
\begin{equation}
\rho^2 = \rho_\mathrm{LISA}^2 + \rho_\mathrm{LGWA}^2 + \rho_\mathrm{XG}^2.
\end{equation}
We classify events with \(\rho \geq 8\) as detectable. {Applying the detection threshold to the network SNR implicitly assumes overlapping observation periods for LISA, LGWA, and XG. However, depending on the astrophysical model,  $88.2$--$99.9\%$ of the events satisfying the network-SNR threshold also exceed $\rho=8$ in at least one individual detector. Our conclusions are therefore largely insensitive to whether or not the three detectors operate contemporaneously.} Figure~\ref{fig:ls_det_mod} shows the merger events from different models detectable by individual detectors (although for our analysis, we apply a single threshold on the network SNR $\rho$). Fig.~\ref{fig:detection_plot} shows the number of detectable mergers from different astrophysical models that we consider. More details are given in Appendix~\ref{sec:sim_detections}. 

For each detected event $i$ we model the expected posterior distributions of source parameters $\bm{\lambda} \equiv {\{\log_{10} M, z, q, \chi_1, \chi_2, \chi^\perp_1, \chi^\perp_2, \cos \iota, \mathrm{RA}, \sin \mathrm{DEC}, \psi, \phi_c, t_{\rm gps}\}}$  (see Table~\ref{tab:priors}) making use of the Fisher information matrix $\Gamma_{jk}$ computed from the GW signal.
\begin{equation}
p(\bm{\lambda} \mid d_i) \propto p(\bm{\lambda} ) ~ p(d_i \mid \bm{\lambda}), 
\label{eq:posterior}
\end{equation}
where $p(\bm{\lambda})$ is the assumed prior distribution of source parameters, while 
\begin{equation}
 p(d_i \mid \bm{\lambda}) \propto \exp \left[- \frac{1}{2}  ~ \Gamma_{jk, i} ~  (\lambda^j - \lambda_{\mathrm{tr}, i}^j) \,   (\lambda^k - \lambda_{\mathrm{tr}, i}^k) \right], 
 \label{eq:likelihood}
\end{equation}
is the Gaussian approximation of the likelihood (see, e.g., \cite{vallisneri}). Here, $\lambda_{\mathrm{tr}, i}^j$ denotes the true value of the parameter $\lambda^j$. Repeated indices are assumed to be summed over.  $\Gamma_{jk, i}$ is the combined Fisher matrix obtained from the $i$th event,
\begin{equation}
\label{eq:fisher_mat}
\Gamma_{jk, i}	 = \Gamma_{jk, i}^{\rm LISA} + \Gamma_{jk, i}^{\rm LGWA} + \Gamma_{jk, i}^{\rm XG}.
\end{equation}

 {To calculate the expected SNR and statistical errors using XG and LGWA observations, we use the \textsc{GWFish} package~\citep{gwfishpackage}, employing the \textsc{IMRPhenomXPHM} waveform family~\citep{IMRPhenomXPHM_Pratten_2021}. This family models the inspiral, merger and ringdown of BBHs including the effect of spin precession. For the same computations in LISA, we use the \textsc{LISAbeta} package~\citep{lisabetapackage}, as it models the antenna-pattern modulations due to the orbital motion of LISA more accurately. Since the \textsc{IMRPhenomXPHM} waveform family is not compatible with \textsc{LISAbeta}, we use the non-precessing waveform family \textsc{IMRPhenomHM}~\citep{IMRPhenomHM_London_2018} for the LISA analysis. Thus, the Fisher matrix components corresponding to the parameters $\chi^\perp_1, \chi^\perp_2$ are set to zero in $\Gamma_{jk, i}^{\rm LISA}$. Essentially, we make the conservative assumption that LISA does not contribute to the measurement of $\chi^\perp_1, \chi^\perp_2$.}

{Note that the Fisher matrix is originally computed in terms of parameters that are more closely related to the waveforms; e.g., the redshifted chirp mass $\mathcal{M}^z_c$ and the luminosity distance $d_L$ instead of $\log_{10} M$ and $z$. To convert $d_L$ into $z$, we assume the $\Lambda$CDM cosmological model with parameters given in~\cite{Planck18_params}. We also include the effect of weak lensing in our posterior samples, which biases our $d_L$ measurement following the prescription in \citet{weak_lensing2}; more details are given in Appendix~\ref{sec:wlensing}. In addition, \textsc{LISAbeta} and \textsc{GWFish} use different parameter conventions: the former employs ecliptic sky coordinates and the solar-system-barycentre arrival time, whereas the latter uses equatorial coordinates and the geocentric GPS time. We transform the original Fisher matrix to the common parameter set ${\bm \lambda}$ using $\Gamma_{\rm trans}=J^{T}\Gamma_{\rm orig}J$, where $J$ is the corresponding Jacobian. The transformed Fisher matrices are then summed to obtain the network Fisher matrix for each event [see Eq.~\eqref{eq:fisher_mat}].}

{After getting the combined transformed Fisher matrix in terms of our required parameters, we compute the likelihood as in Eq.~\eqref{eq:likelihood}. Using this likelihood and the priors on those source parameters $\bm{\lambda}$ given in Table~\ref{tab:priors}, we sample the posterior distribution [see Eq.~\eqref{eq:posterior}]. Since we use priors that are flat in $\bm \lambda$,  no reweighting on individual events is required. An example of the full posterior distribution is shown in Fig.~\ref{fig:full_corner}. We further marginalize this posterior distribution (in the 13-dimensional parameter space) over all other parameters to compute the posterior distribution in $\log_{10} M$ and $z$. An example of the  marginalized distribution in $\log_{10} M$ and $z$ is shown in Fig.~\ref{fig:2d_corner_post_recon}.}

From these synthetic posterior distributions of detectable mergers, we reconstruct the mass-redshift distribution of the astrophysical population using the nonparametric Bayesian inference code \textsc{Figaro}~\citep{figaropackage, figaro2024}. It uses Dirichlet Process Gaussian Mixture Models (DPGMM) and their hierarchical extension (HDPGMM). This method aims to model potentially complex distributions as Gaussian mixtures, with both the number of components and their parameters inferred directly from the data (see Fig.~\ref{fig:2d_corner_post_recon}). HDPGMM also takes into account the detector selection effects through the detection probability \(p_{\rm det}(\bm{\lambda})\). Using the posterior samples from the detected sources together with the detector selection function, it reconstructs the posterior predictive distributions (PPDs) of the intrinsic population (see Fig.~\ref{fig:hdpgmm_recon_popIII}), which represent the allowed realizations of the intrinsic mass-redshift distribution compatible with the observed data. Using these PPDs and the observed number of mergers, we reconstruct the posterior on the merger-rate density (see Appendix~\ref{sec:merger_rate_recons}
for details). 

To ensure a stable and robust reconstruction, we impose a cut on the detection probability and only retain those posterior samples from each of the detected events which satisfy \(p_{\rm det} > 0.1\). This threshold is applied to avoid the contribution from sources with extremely small detection probabilities. Such events are strongly affected by selection effects and receive very large weights when correcting for detector sensitivity, which can artificially dominate the reconstruction and introduce significant statistical fluctuations or biases in the inferred population distribution.
\section{Results}
\label{sec:results}

\begin{figure*}[tbh]
\centering	
\includegraphics[width=0.7\columnwidth]{plots/popIII_ind_reconst_plot.pdf}
\includegraphics[width=0.7\columnwidth]{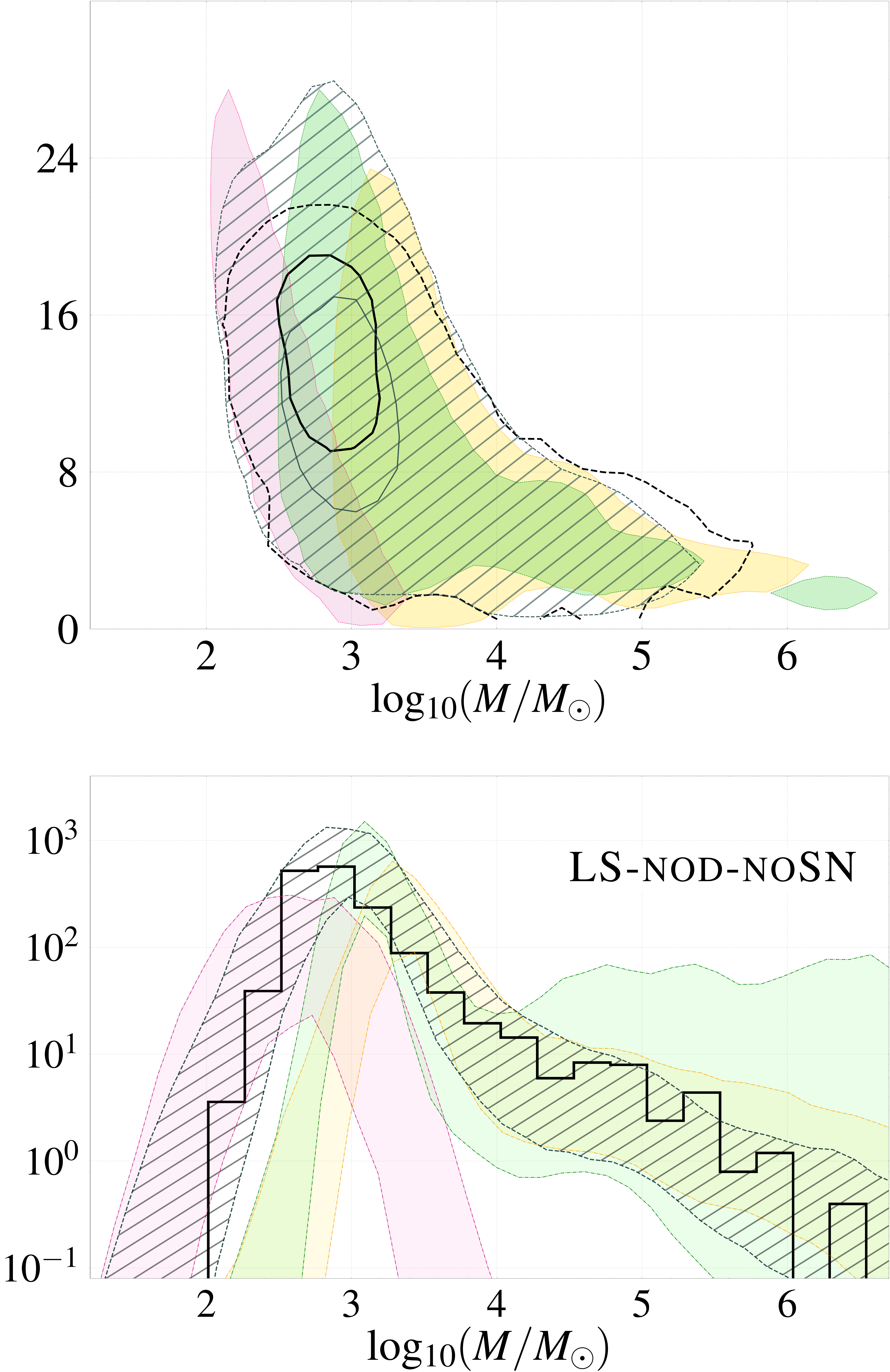}
\includegraphics[width=0.7\columnwidth]{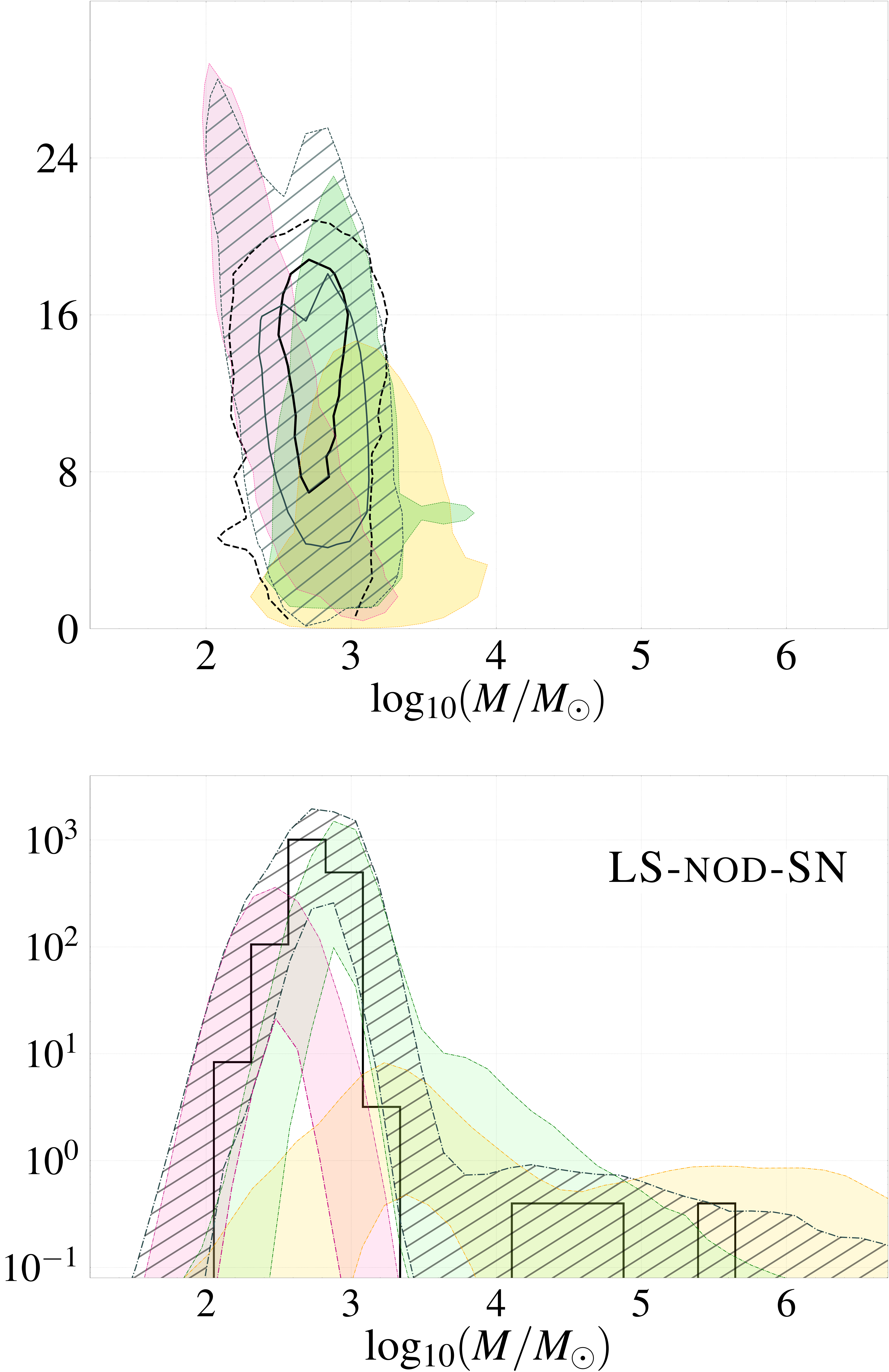}
\caption{\textit{Top}: Reconstructed mass-redshift distribution of BBH mergers from the LS models, by the LISA, LGWA and XG detectors as well as from multiband observations. XG, LGWA and LISA are better at reconstructing the low-, intermediate-, and high-mass regions of parameter space, respectively. The reconstruction using multiband observations (gray contours) is more accurate than that obtained with any individual detector. \textit{Bottom}: Reconstructed mass distribution marginalized over the redshift.}
\label{fig:ls_lisa_vs_combined}
\end{figure*}

\begin{figure*}[tbh]
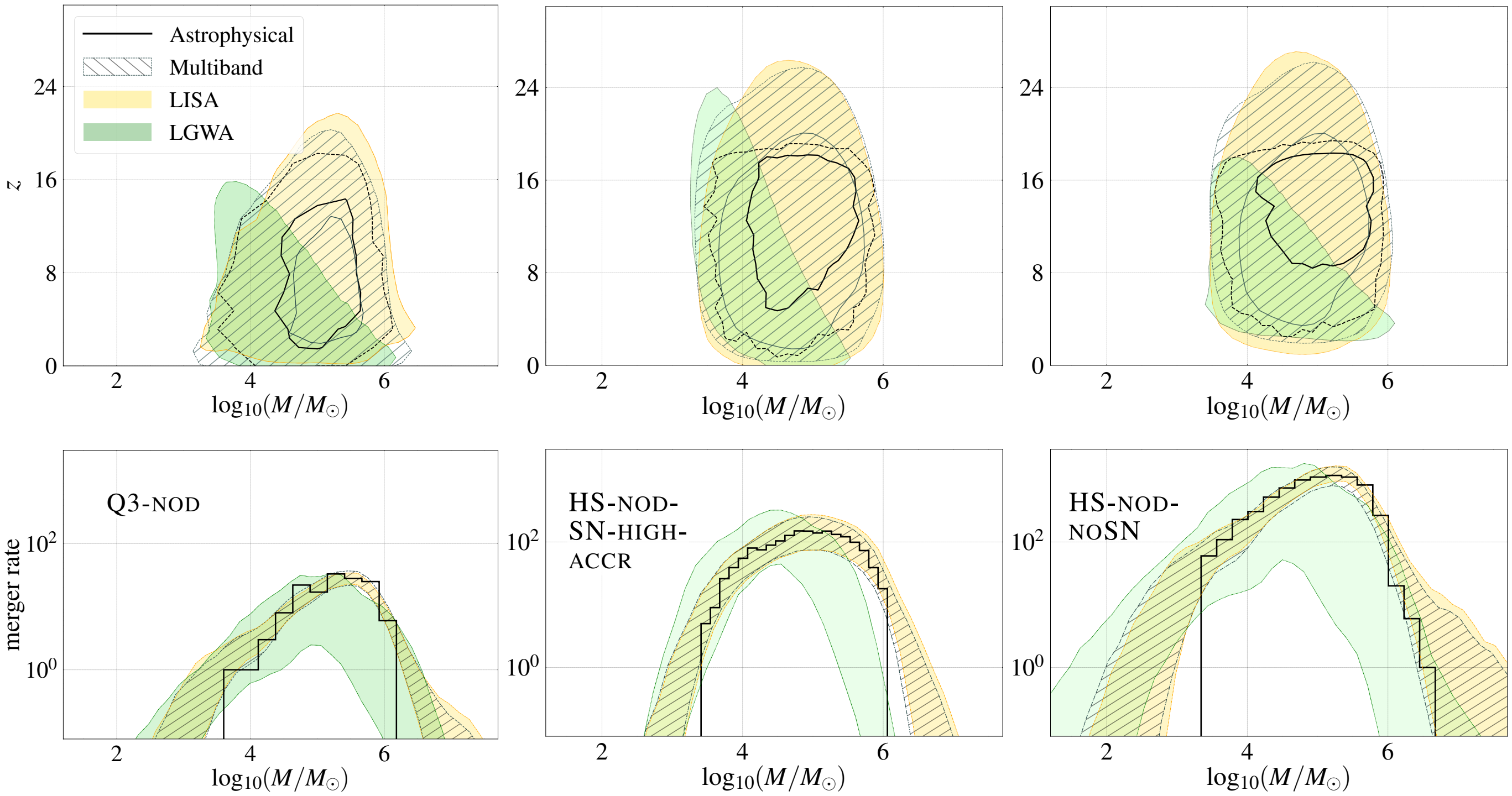

\centering	
\includegraphics[width=0.7\columnwidth]{plots/q3nod_ind_reconst_plot.pdf}
\includegraphics[width=0.7\columnwidth]{plots/hs_high_accr_ind_reconst_plot.pdf}
\includegraphics[width=0.7\columnwidth]{plots/hs_noSN_ind_reconst_plot.pdf}
\caption{Same as Fig.~\ref{fig:ls_lisa_vs_combined} except that here we consider the HS models, and the reconstruction is shown only for LISA and LGWA detectors. For the HS models considered, LISA, by itself, does a good job of reconstructing the mass-redshift distribution.}
\label{fig:hs_lisa_vs_comb}
\end{figure*}

{Figure~\ref{fig:ls_recons_plots} presents the reconstructed observer-frame merger-rate density, ${d^3N}/({dt_{\rm o}\,d\log_{10}M\,dz})$, for the LS models assuming a 10-year observing period. To assess the reconstruction accuracy, we overlay the true injected population and contours of the detection probability for each of the three detectors considered.\footnote{Because we retain only posterior samples from regions with $p_{\rm det}>0.1$, the reconstructed mass-redshift distribution is expected to be largely confined to those regions, whereas the true distribution is not subjected to this cut. In some cases, such as \textsc{LS-nod-noSN} and \textsc{LS-nod-SN} in Fig.~\ref{fig:ls_recons_plots}, the reconstructed distribution extends slightly into regions with $p_{\rm det}<0.1$ because of smoothing by the DPGMM.} Figures~\ref{fig:ls_recons_plots_4yrs} and \ref{fig:ls_recons_plots_1yrs} show the corresponding reconstructions for observing periods of 4 years and 1 year, respectively. In all three cases, the reconstructed distributions broadly recover the true populations, with the reconstruction becoming more precise as the observing time increases.}

{Figures~\ref{fig:hs_recons_plots} and \ref{fig:hs_recons_plots_1yr} present the corresponding reconstructions for the HS models assuming observing periods of 10 years and 1 year, respectively. Even in the 10-year observing scenario, LISA retains its nominal 4-year mission duration and dominates the reconstruction of the HS populations. We therefore do not present a separate 4-year reconstruction for these models.}

In Fig.~\ref{fig:ls_lisa_vs_combined} and Fig.~\ref{fig:hs_lisa_vs_comb}, we compare the reconstruction by individual detectors with that using the full multiband detector network. For the LS models, the multiband reconstruction significantly outperforms the reconstruction using any single detector. This is because the astrophysical model predicts BH mergers across a wide range of masses and redshifts, which are not completely probed by any single detector. An interesting feature of the \textsc{LS-nod-SN} model is observed in the reconstruction: since a significant fraction of the true mergers of this model falls in the blind region of the detector network (between $2.5 < \log_{10} M < 3$ and $z\sim{15}$), we are mildly underestimating the true merger rate of this model at high z ($\sim{15}$) (Fig.~\ref{fig:ls_recons_plots}). In contrast, for the HS models, LISA alone is already capable of reconstructing almost the entire population distribution (Fig.~\ref{fig:hs_lisa_vs_comb}). The reason is that the HS population produces most of its mergers in the LISA band. Although we get some mergers in the low mass end falling in the LGWA  band, these events are mostly multiband events whose SNRs are dominated by LISA. {Note that, for the HS models, both the ``true'' and the reconstructed distributions refer to the $q\geq0.1$ subpopulation only.}

\begin{figure*}[t]
\centering	
\includegraphics[width=1.8\columnwidth]{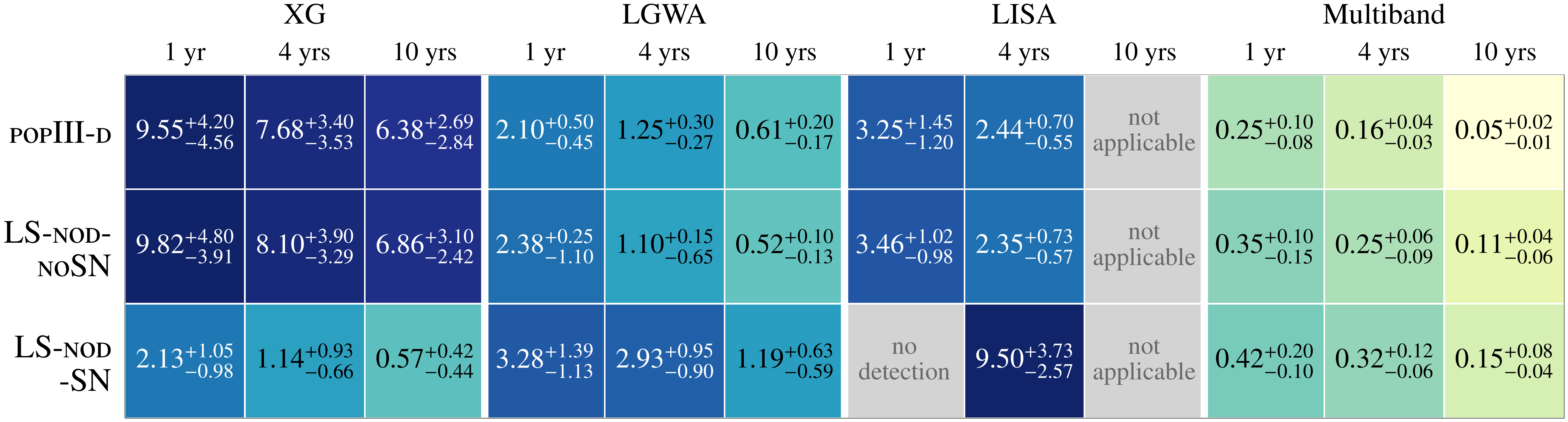}
\caption{KL divergence values $D_{\rm KL}$ between the true $(\log_{10}M, z)$ distribution of the LS models and their reconstructions from the events detected by a single detector, or multiband detections. Central values correspond to the median reconstruction, and errors show the 1$\sigma$ range over the reconstruction draws. Lower values (lighter colors) mean a reconstruction closer to the truth and the multiband reconstruction gives the lowest $D_{\rm KL}$ for every model. The large value for \textsc{LS-nod-SN} with LISA  comes from the very small number of LISA detections in this model and hence the reconstruction being poorly constrained.}
\label{fig:dkl_multibanding}
\end{figure*}

\begin{figure}[tbh]
\centering	
\includegraphics[width=1\columnwidth]{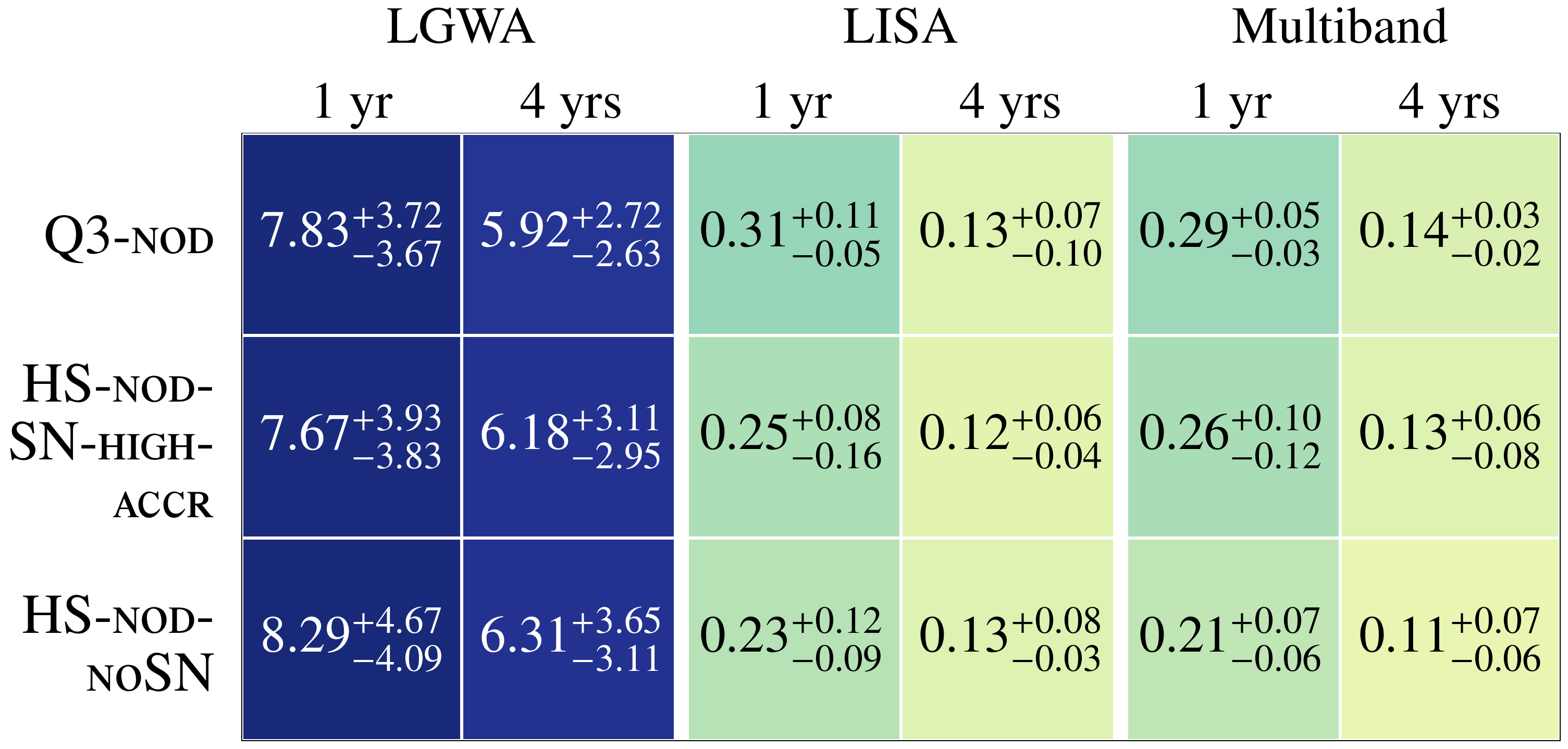}
\caption{Same as Fig.~\ref{fig:dkl_multibanding}, except that this figure corresponds to the HS models. There are no XG detections; The LISA and multiband results agree within statistical fluctuations.}
\label{fig:dkl_multibanding_HS}
\end{figure}

To quantify the discrepancy of the reconstructed mass-redshift distribution ${p_{\rm rec}(\log_{10}M,z)}$ with the true distribution $p_{\rm tr}(\log_{10}M,z)$, we compute the Kullback-Leibler (KL) divergence~\citep{kullback1951information} between the two:
\begin{equation}
D_{\rm KL} = \int p_{\rm tr}(\log_{10}M,z) \ln\!\left[\frac{p_{\rm tr}(\log_{10}M,z)}{p_{\rm rec}(\log_{10}M,z)} \right] \,d\log_{10}M\,dz.
\label{eq:kl_divergence}
\end{equation}
Both distributions are normalized to unity before evaluating Eq.~\eqref{eq:kl_divergence}. Thus, $D_{\rm KL}$ quantifies differences in the shape of the distributions, rather than differences in their overall merger rates. Values close to zero suggest excellent reconstruction, whereas larger values indicate poorer reconstruction. We evaluate $D_{\rm KL}$ separately for each PPD draw and report the median and the corresponding $1\sigma$ interval over these draws (Fig.~\ref{fig:dkl_multibanding} for LS models and Fig.~\ref{fig:dkl_multibanding_HS} for HS models). The KL divergence values are consistent with the reconstructed mass-redshift distributions.  For example, in the reconstruction of the \textsc{popIII-d} model, LGWA makes the most significant contribution since most of the mergers are in the mass-redshift space accessible to LGWA; the next significant role is played by LISA, while XG makes only a modest contribution. In contrast, for the \textsc{LS-nod-SN} model, the most significant contributions are from XG and LGWA. On the other hand, the reconstruction of all the HS models is entirely dominated by LISA. 

\section{Summary and future work}
\label{sec:summary}

The formation and growth of SMBHs remain among the central unsolved problems in astrophysics. Major uncertainties persist regarding the initial mass function of BH seeds, the nature and efficiency of accretion, the co-evolution of galaxies and massive BHs, and the roles of SN and AGN feedback. Over the coming decade, electromagnetic observations from facilities such as Roman, Rubin, Euclid, Athena, and SKA will substantially improve constraints on SMBH demographics by probing quasars, AGNs, tidal disruption events, and other tracers over a broad range of masses and redshifts. At the same time, GW observations in the millihertz, decihertz, and audio-frequency bands will provide an independent and complementary view of SMBH assembly through the merger history of massive BH binaries. In this work, we have investigated the ability of future multiband GW observations to reconstruct the mass-redshift distribution of merging BH binaries across a wide range of masses and cosmic epochs. {For our fiducial analysis, we assume overlapping observing periods of 4 years for LISA and 10 years for LGWA and XG. Our conclusions are nevertheless expected to remain broadly unchanged if the observing periods do not overlap, since most events in the multiband sample exceed the detection threshold in at least one individual detector. To examine the dependence of the reconstruction on observing time, we also consider shorter scenarios in which all detectors operate for 4 years or 1 year.}

Our results demonstrate that multiband observations significantly enhance the reconstruction of the merger population in scenarios involving light BH seeds. In such models, mergers populate a broad region of the mass-redshift plane, making it difficult for any single detector to provide complete coverage. The combination of space- and ground-based observatories therefore yields substantially greater constraining power than any individual instrument. {Even 1 year of multiband observations is significantly better than 10 years of observations using any single detector.} In contrast, for HS models, where the merger population is concentrated at higher masses, LISA alone is expected to detect the majority of events and recover much of the available information. 

{Several limitations of this first demonstration should be kept in mind. We account for parameter-estimation uncertainties using approximate event-level posteriors, but neglect noise-induced fluctuations by centering each posterior on the injected parameters, apart from the applied weak-lensing shift. Moreover, our selection function in the $(M,z)$ plane is obtained by marginalizing over the model-dependent distributions of mass ratio and BH spins, and therefore assumes knowledge of the underlying population. A fully self-consistent and model-independent analysis should instead reconstruct the joint distribution of the relevant intrinsic parameters, including $(M,q,z,\chi_{1},\chi_{2},\chi_{1}^{\perp},\chi_{2}^{\perp})$, while evaluating the (model-independent) selection function directly in this higher-dimensional space. Our forecasts also rely on Fisher-matrix likelihoods, idealized detector sensitivities and observing schedules, a fixed cosmology, and simplified treatments of weak lensing. Future work should address these limitations through full Bayesian parameter estimation, realistic detector duty cycles and mission overlap, waveform and calibration uncertainties, improved treatments of lensing and selection effects, and a broader range of astrophysical populations. Extending the reconstruction to include mass ratios and spins, and combining GW measurements with electromagnetic observations of high-redshift AGNs and quasars, should provide stronger constraints on the physical processes governing SMBH seeding, accretion, and merger-driven growth.}

\paragraph{Acknowledgements:} We thank Ankur Barsode, Rajaram Nityananda, Preeti Kharb, Prayush Kumar and other members of the ICTS Astrophysical Relativity group for useful discussions. We acknowledge support of the Department of Atomic Energy, Government of India, under projects No. RTI4019 and No. RTI4013. PA and AK also acknowledge support from the ICTP through the Associates Programme and from the Simons Foundation through the grant (Record ID: SFI-MPS-T-Institutes-00012057, AD). SJ acknowledges grants from the Research Grants Council of Hong Kong (Project No. CUHK 14304622, 14307923 and 14307724) as well as the Postdoctoral Fellowship Scheme. EB acknowledges support from the European Union's Horizon ERC Synergy Grant ``Making Sense of the Unexpected in the Gravitational-Wave Sky'' (Grant No.~GWSky-101167314) and from the Agenzia Spaziale Italiana (ASI), Project n.~2024-36-HH.0, ``Attivit\`a per la fase B2/C della missione LISA''. This work makes use of the \texttt{numpy}~\citep{harris2020array}, \texttt{scipy}~\citep{2020SciPy-NMeth},  \texttt{astropy}~\citep{astropy:2013, astropy:2018, astropy:2022}, \texttt{matplotlib}~\citep{Hunter:2007}, \texttt{HTcondor}~\citep{Thain:2005}, and \texttt{pyCBC}~\citep{alex_nitz_2024_10473621},
\texttt{lisabeta}~\citep{lisabetapackage}, \texttt{GWFish}~\citep{gwfishpackage}, \texttt{Figaro}~\citep{figaropackage, figaro2024} software packages. Numerical simulations were performed on the Alice computing cluster at the International Centre for Theoretical Sciences, Tata Institute of Fundamental Research.

\appendix
\section{Simulating BBH mergers}
\label{sec:sim_mergers}

\subsection{Generating BBH samples from an astrophysical population model}

Here we describe how a weighted population-synthesis catalog from \citet{Barausse_2012} is converted into a Monte Carlo realization of the observer-frame BH merger population. The observer-frame merger rate distribution ${d^2N}/{dt_{\rm o}\,dz}$ is
related to the source-frame comoving merger-rate density $n(z) \equiv {d^2N}/{dt_{\rm s}\,dV_c}$ by
\begin{equation}
 \frac{d^2N}{dt_{\rm o}\,dz} = n(z) ~ \frac{dV_c}{dz} ~ \frac{1}{1+z}
\label{eq:observer_merger_rate}
\end{equation}
where ${dV_c}/{dz}$ is the differential comoving volume and $1/(1+z)$ accounts for cosmological time dilation.

Each binary $i$ in the population-synthesis catalog carries a statistical weight $W_i$, equal to the comoving number density of its host halo as given by the halo mass function at $z=0$, divided by the number of merger-tree realizations run for that halo mass. Because comoving number densities are conserved along a merger tree, each entry stands for $W_i$ mergers per unit comoving volume taking place at redshift $z_i$. If the catalog is binned in redshift intervals of width $\Delta z$, each bin spans a source-frame time $\Delta t_{\rm s}=\Delta z/[(1+z)H(z)]$, so that the comoving merger-rate density is
\begin{equation}
n(z) = \frac{(1+z)\,H(z)}{\Delta z}\sum_{i\in{\rm bin}} W_i.
\label{eq:nz_catalog}
\end{equation}
Inserting Eq.~\eqref{eq:nz_catalog} into Eq.~\eqref{eq:observer_merger_rate} and using $dV_c/dz=4\pi c\,D_c^2(z)/H(z)$, where $D_c$ is the comoving distance, the factors of $(1+z)$ and $H(z)$ cancel and we obtain
\begin{equation}
\frac{d^2N}{dz\,dt_{\rm o}} = \frac{4\pi c}{\Delta z}\sum_{i\in{\rm bin}} W_i\,D_c^2(z_i).
\label{eq:d2N_catalog_final}
\end{equation}
The expected number of mergers in an observing period $T_{\rm obs}$ follows by integrating over redshift and observer-frame time, $\Lambda = T_{\rm obs} \mathcal{R} = T_{\rm obs}  \int dz ~ d^2N/(dz\,dt_{\rm o})=\sum_i\Lambda_i$, where the sum now runs over the whole catalog, the bin width drops out, and
\begin{equation}
\Lambda_i = 4\pi c\,D_c^2(z_i)\,W_i\,T_{\rm obs}
\label{eq:lambda_def}
\end{equation}
is the mean number of mergers contributed by binary $i$ during $T_{\rm obs}$. Above, $\mathcal{R} \equiv dN/dt_{\rm o} = \int dz ~ d^2N/(dz\,dt_{\rm o})$ is the merger rate in the observer frame. Treating mergers as independent events, we include each catalog binary $N_i$ times, drawn from the Poisson distribution 
\begin{equation}
p(N_i) = \frac{\Lambda_i^{N_i} e^{-\Lambda_i}}{N_i!}. 
\label{eq:poisson_draw}
\end{equation}
Applying this to every binary yields one Monte Carlo realization of the merger population that the astrophysical model predicts for an observing period $T_{\rm obs}=10$\,yr.

\subsection{Computing SNRs in different detectors}
\label{sec:sim_detections}

\begin{table}[tbh]
  \centering
  \begin{tabular}{lcc}
    \toprule
    Model & $N_{\rm det}$ & $N_{\rm inspiral}$ \\
    \midrule
    \multicolumn{3}{l}{\textit{Light seeds}} \\
    \textsc{popIII-d}       & 338   & 10 \\
    \textsc{LS-nod-noSN}    & 232   & 15 \\
    \textsc{LS-nod-SN}      & 15    &  2 \\
    \midrule
    \multicolumn{3}{l}{\textit{Heavy seeds}} \\
    \textsc{Q3-nod}                  & 540   & 3 \\
    \textsc{HS-nod-SN-high-accr}     & 6145  & 4 \\
    \textsc{HS-nod-noSN}             & 29387 & 4 \\
    \bottomrule
  \end{tabular}
    \caption{Number of LISA detections during a four-year mission.
  $N_{\rm det}$ is the number of detectable binaries that merge during
  the mission, while $N_{\rm inspiral}$ is the number detectable by
  LISA that merge after the mission and subsequently enter the LGWA or
  XG band. We exclude these inspiral-only events from our analysis.
  Light-seed models generally predict more such events because their
  lower-mass binaries evolve more slowly through the LISA band. The numbers here are after the $q \geq 0.1$ cut that we have imposed on the population.}
  \label{tab:ndet}
\end{table}

We identify the binaries observable by each detector using the following procedure:

\begin{enumerate}

    \item We adopt a nominal observer-frame time window of $T_{\rm obs}=10~{\rm yr}$, during which the population model predicts $N$ BBH mergers. We assign each binary a coalescence time $t_c$ drawn uniformly from $\mathcal{U}(t_0,t_0+T_{\rm obs})$, where $t_0$ is the beginning of the observing window. Equivalently, the time from $t_0$ to coalescence, $\tau\equiv t_c-t_0$, is distributed as $\mathcal{U}(0,T_{\rm obs})$.

    \item For a binary with time to coalescence $\tau$, the GW frequency at the beginning of the observing window is, at leading post-Newtonian order,
    \begin{equation}
        f_{\min}(\tau,\mathcal{M}_{c}^z) =  \frac{1}{\pi} \left(\frac{5}{256\tau}\right)^{3/8} \left(\frac{G\mathcal{M}_{c}^z}{c^3}\right)^{-5/8},
        \label{eq:fmin_time_to_merger}
    \end{equation}
    where $\mathcal{M}_c^z \equiv (1+z)\mathcal{M}_c$ is the detector-frame chirp mass. The frequency $f_{\min}$ is further restricted to the detector's sensitive frequency band.

    \item While LGWA and XG detectors are likely to have observing periods of over a decade, LISA is expected to have a shorter mission period, $T_{\rm obs}^{\rm LISA} \simeq 4~{\rm yr}$. Because the coalescence times are distributed uniformly over the 10-year window, a fraction $T_{\rm obs}^{\rm LISA}/T_{\rm obs} \simeq 0.4$ of the binaries are expected to merge during the LISA mission. Binaries with $\tau>T_{\rm obs}^{\rm LISA}$ do not merge during the mission but may be observed as \emph{inspiralling} binaries. Their frequency at the end of the mission is 
    \begin{equation}
        f_{\max} =  f_{\min}\!\left(\tau-T_{\rm obs}^{\rm LISA},\mathcal{M}_c^z \right),  \qquad \tau>T_{\rm obs}^{\rm LISA}.
        \label{eq:fmax_lisa_inspiral}
    \end{equation}
    For binaries that merge during the mission ($\tau\leq T_{\rm obs}^{\rm LISA}$), we initially set $f_{\max}=f_{\rm cut}$, where $f_{\rm cut}$ is the upper cut-off frequency of a non-spinning BBH signal~\citep{fcutphenomd}.

    \item For each detector, we compute the optimal SNR as
    \begin{equation}
        \rho^2=\langle h\mid h\rangle =  4\,{\rm Re}\int_{f_{\min}}^{f_{\max}} \frac{\tilde{h}^{*}(f)\tilde{h}(f)} {S_n(f)}\,df,
        \label{eq:optimal_snr}
    \end{equation}
    where $S_n(f)$ is the detector noise power spectral density. In all cases, the upper integration limit is restricted to
    \begin{equation}
        f_{\max}\rightarrow\min(f_{\max},f_{\rm cut}),
    \end{equation}
    and both integration limits are restricted to the detector's sensitive frequency band.  In our analysis, we find that the number of \emph{inspiralling} binaries is much smaller than the number of \emph{merging} binaries (see Table~\ref{tab:ndet}). We do not include these inspiral-only events in the subsequent analysis.
    \end{enumerate}

\subsection{Effect of weak lensing}
\label{sec:wlensing}

\begin{figure*}[tbh]
\centering	
\includegraphics[width=1\columnwidth]{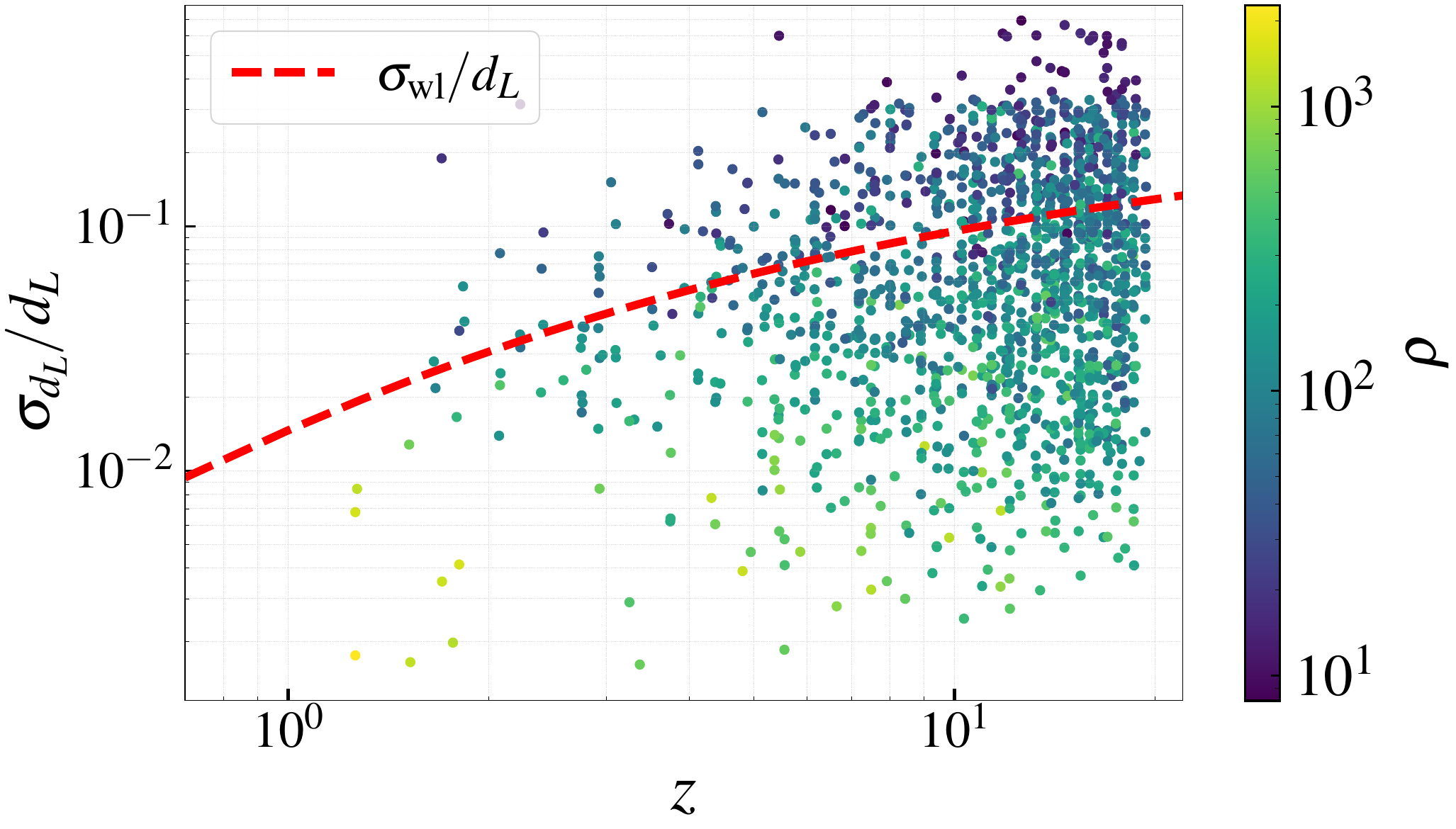}	
\hspace{0.02\columnwidth}
\includegraphics[width=1\columnwidth]{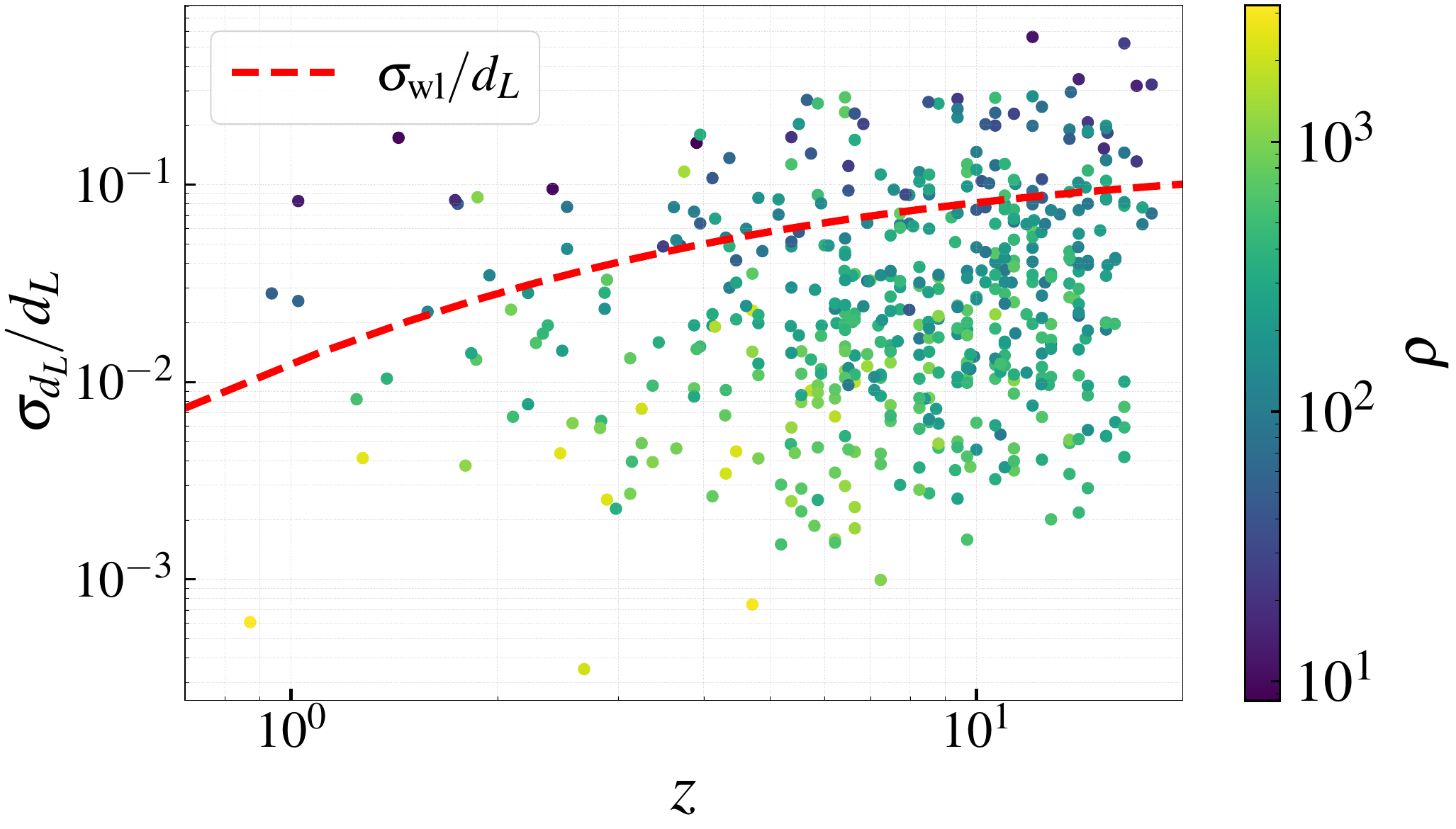}
\caption{The red dashed line shows the weak lensing 1$\sigma$ relative error for one of the LS models (left) and one of the HS models (right). The scatter plot shows the Fisher errors for the detected population for these two models with color denoting the network SNR $\rho$. For high-SNR signals, the weak lensing error could dominate the error budget.}
\label{fig:weak_lensing}
\end{figure*}

Weak lensing arises from the gravitational deflection of GWs by the inhomogeneous large-scale structure along the path between the source and the observer. This effect introduces a small magnification/demagnification of the observed signals, and consequently an additional error $\delta d_L$ in the recovered luminosity distance. Following the prescription of \cite{weak_lensing2}, for each event located at a redshift $z$, we model $\delta d_L$ as a Gaussian random variable with mean zero and standard deviation: 
\begin{equation}
    {\sigma_{\text{wl}}(z)} = {d_L (z)} ~ 0.066 \left( \frac{1 - (1+z)^{-0.25}}{0.25} \right)^{1.8}.
    \label{eq:wl}
\end{equation}
As done in \cite{weak_lensing2}, we also consider the optimistic scenario and divide the $\sigma_{\text{wl} (z)}$ by a factor of two. This shift is then added to the true luminosity distance as $d_L^{\text{obs}} = d_L^{\text{true}} + \delta d_L$. We then generate posterior samples of $d_L$ for all the events using $d_L^{\text{obs}}$ as the mean and $\sigma_{d_L}^2$, calculated from the Fisher matrix, as the variance of our approximate posteriors. We show the weak-lensing error and corresponding Fisher errors for a typical LS (left) and HS model (right) in Fig.~\ref{fig:weak_lensing}.

Because we shift the posterior mean by $\delta d_L$ while retaining only the Fisher variance, the inferred luminosity distance of a sufficiently loud event may be significantly biased. These event-level biases propagate into the reconstructed mass-redshift distribution and are therefore already present in our results. A more complete analysis would marginalize over the unknown lensing magnification, for example by adding $\sigma_{\rm wl}^2$ in quadrature to the Fisher variance of each event, as in \citet{weak_lensing2}. This would broaden the event-level posteriors and reduce the influence of lensing-induced shifts on the population reconstruction. Our present treatment should be regarded as an approximate assessment of the effect of unmodeled weak lensing in the reconstruction of the mass-redshift distribution.

\subsection{Selection effects}
\label{sec:sel_effects}

The probability of detecting a GW signal depends on its intrinsic and extrinsic parameters. For example, nearby and more massive binaries generally produce larger SNRs and are therefore more likely to be detected. Accounting for this selection effect is essential for reconstructing the underlying population without biasing the inferred mass and redshift distributions toward louder sources. We estimate the selection function as follows:

\begin{enumerate}
    \item We construct a three-dimensional grid in $(\log_{10}M,q,z)$ over the ranges $0\leq\log_{10}(M/M_\odot)\leq10, 0.1\leq q\leq1, 10^{-4}\leq z\leq50$. At each grid point, we generate $N_{\rm inj}\simeq5000$ simulated signals and marginalize over the remaining intrinsic and extrinsic parameters. {The spin parameters $\chi_1, \chi_2, \chi^\perp_1, \chi^\perp_2$ are sampled from the corresponding spin distribution $p_{\rm pop} (\chi_1, \chi_2, \chi^\perp_1, \chi^\perp_2)$ of the model. The remaining parameters are drawn from the distributions listed in Table~\ref{tab:priors}.} 
    Coalescence times are drawn uniformly over the nominal 10-year observing window. Because we include only binaries that merge during the four-year LISA mission, we set $\rho_{\rm LISA}=0$ for systems with coalescence times later than the end of that mission.

    \item For every simulated signal, we compute the network SNR by combining the contributions from LISA, LGWA, and XG in quadrature:
    \begin{equation}
        \rho_{\rm net}^2(\log_{10}M,q,z)=\sum_{D}\rho_D^2(\log_{10}M,q,z),
        \label{eq:network_snr_selection}
    \end{equation}
    where ${D\in\{\mathrm{LISA,LGWA,XG}\}}$ and the dependence on the marginalized parameters is implicit.

    \item We classify an injection as detectable if $\rho_{\rm net}\geq8$. The selection function at each grid point is then estimated as
    \begin{equation}
        p_{\rm det}(\log_{10}M,q,z)=\frac{N_{\rm det}(\log_{10}M,q,z)}{N_{\rm inj}(\log_{10}M,q,z)},
        \label{eq:selection_function_3d}
    \end{equation}
    where $N_{\rm det}$ is the number of injections that satisfy the detection criterion.
\end{enumerate}

Because the population reconstruction is performed in $(\log_{10}M,z)$, we marginalize the three-dimensional selection function over the model-dependent mass-ratio distribution:
\begin{equation}
    p_{\rm det}(\log_{10}M,z)=\int p_{\rm det}(\log_{10}M,q,z)\,p_{\rm pop}(q\mid\log_{10}M,z)\,dq.
    \label{eq:selection_function_2d}
\end{equation}
The resulting $p_{\rm det}(\log_{10}M,z)$ gives the probability that a binary with source-frame total mass $M$ and redshift $z$ is detected after marginalizing over its mass ratio and the remaining binary parameters.

\section{Inferring the observer-frame merger-rate density using \textsc{Figaro}}
\label{sec:merger_rate_recons}

Let $p_{\rm pop}(\boldsymbol{\theta})$ denote the normalized distribution of the intrinsic population, where $\boldsymbol{\theta}=(\log_{10}M,z)$. The fraction of this population that is detectable is
\begin{equation}
    \alpha \equiv \int p_{\rm det}(\boldsymbol{\theta})\,p_{\rm pop}(\boldsymbol{\theta})\,d\boldsymbol{\theta}, \qquad \int p_{\rm pop}(\boldsymbol{\theta})\,d\boldsymbol{\theta}=1.
    \label{eq:detection_efficiency}
\end{equation}
Here, $p_{\rm det}(\boldsymbol{\theta})$ is the probability of detecting an event with parameters $\boldsymbol{\theta}$.

\textsc{Figaro} reconstructs the intrinsic population through a set of normalized PPD draws,
\begin{equation}
    \int p_{\rm ppd}^{(j)}(\boldsymbol{\theta})\,d\boldsymbol{\theta}=1, \qquad j=1,\ldots,N_{\rm PPD}.
\end{equation}
Each PPD draw represents a possible realization of $p_{\rm pop}(\boldsymbol{\theta})$ and therefore implies a corresponding detection efficiency
\begin{equation}
    \alpha_j = \int p_{\rm det}(\boldsymbol{\theta})\,p_{\rm ppd}^{(j)}(\boldsymbol{\theta})\,d\boldsymbol{\theta}.
    \label{eq:ppd_detection_efficiency}
\end{equation}

Let $\mathcal{R}\equiv dN/dt_{\rm o}$ denote the total observer-frame merger rate, and let $T_{\rm obs}$ be the observing time. For a given population and its detection efficiency $\alpha$, the expected number of detected events is
\begin{equation}
    \Lambda_{\rm det}=\mathcal{R}\,T_{\rm obs}\,\alpha.
\end{equation}
The likelihood of observing $N_{\rm det}$ events is therefore
\begin{equation}
    p(N_{\rm det}\mid\mathcal{R},\alpha)=\frac{\left(\mathcal{R}T_{\rm obs}\alpha\right)^{N_{\rm det}}e^{-\mathcal{R}T_{\rm obs}\alpha}}{N_{\rm det}!}.
\end{equation}
For a prior $p(\mathcal{R})$, the corresponding posterior on the merger rate follows from Bayes' theorem: 
\begin{equation}
    p(\mathcal{R}\mid N_{\rm det},\alpha)\propto p(\mathcal{R})\left(\mathcal{R}T_{\rm obs}\alpha\right)^{N_{\rm det}}e^{-\mathcal{R}T_{\rm obs}\alpha}.
    \label{eq:rate_posterior}
\end{equation}
We evaluate Eq.~\eqref{eq:rate_posterior} on a grid in $\mathcal{R}$ and normalize it numerically.

For each PPD draw $j$, we substitute the corresponding $\alpha_j$ into Eq.~\eqref{eq:rate_posterior} to obtain the conditional rate posterior
\begin{equation}
    p_j(\mathcal{R}\mid\{d_i\},N_{\rm det})\equiv p(\mathcal{R}\mid N_{\rm det},\alpha_j).
    \label{eq:conditional_rate_posterior}
\end{equation}
We then draw a rate $\mathcal{R}_j$ from each conditional posterior. The collection of samples $\{\mathcal{R}_j\}$ propagates the uncertainty in the reconstructed population, and hence in the detection efficiency, into the inferred merger rate.

The differential observer-frame merger-rate density associated with the joint draw $\{\mathcal{R}_j,p_{\rm ppd}^{(j)}(\boldsymbol{\theta})\}$ is then estimated from the samples 
\begin{equation}
    \left[\frac{d^3N}{dt_{\rm o}\,d\log_{10}M\,dz}\right]_j=\mathcal{R}_j\,p_{\rm ppd}^{(j)}(\log_{10}M,z).
    \label{eq:differential_rate_draw}
\end{equation}

\bibliographystyle{aasjournalv7}
\bibliography{references}

\end{document}